\documentclass[pdflatex,sn-nature]{sn-jnl}
\usepackage{graphicx}%
\usepackage{multirow}%
\usepackage{amsmath,amssymb,amsfonts}%
\usepackage{amsthm}%
\usepackage{mathrsfs}%
\usepackage[title]{appendix}%
\usepackage{xcolor}%
\usepackage{textcomp}%
\usepackage{manyfoot}%
\usepackage{booktabs}%
\usepackage{algorithm}%
\usepackage{algorithmicx}%
\usepackage{algpseudocode}%
\usepackage{listings}%

\theoremstyle{thmstyleone}%

\theoremstyle{thmstyletwo}%

\theoremstyle{thmstylethree}%

\begin{document}

\title[Article Title]{Clouds as the driver of variability and colour changes in brown dwarf atmospheres}

\author*[1,2]{\fnm{Lucas} \sur{Teinturier}}\email{lucas.teinturier@obspm.fr}

\author[1]{\fnm{Benjamin} \sur{Charnay}}\email{benjamin.charnay@obspm.fr}

\author[2]{\fnm{Aymeric} \sur{Spiga}}\email{aymeric.spiga@lmd.ipsl.fr}
\author[1]{\fnm{Bruno} \sur{Bézard}}\email{bruno.bezard@obspm.fr}
\affil*[1]{\orgdiv{LIRA}, \orgname{Observatoire de Paris, Université PSL, CNRS, Sorbonne Université, Université Paris Cité}, \orgaddress{\street{5 Place Jules Janssen}, \city{Meudon}, \postcode{92195}, \country{France}}}

\affil[2]{\orgdiv{Laboratoire de Météorologie Dynamique}, \orgname{IPSL, CNRS, Sorbonne Université, Ecole Normale Supérieure, Université PSL, Ecole Polytechnique, Institut Polytechnique de Paris}, \orgaddress{\street{4 Place Jussieu}, \city{Paris}, \postcode{75005}, \country{France}}}


\abstract{\textbf{ Brown dwarfs are massive, giant exoplanet analogues subject to variability and colour changes, known as the L/T transition, 
fundamental for their thermal evolution.
The drivers of the L/T transition remain elusive, with atmospheric circulations and/or clouds usually suggested as potential mechanisms.
Using a three-dimensional Global Climate Model including cloud formation, transport and multi-wavelength radiative effects, we show that clouds play a major role in shaping the atmospheric properties of brown dwarfs. 
Cloud radiative effect, which triggers atmospheric convection, leads to spectral, spatial, and temporal variability in the modelled brown dwarfs, in agreement with the observed variability and  L/T transition.
Low latitudes are subject to sustained wave activity, whereas eddies dominate higher latitudes. Our results highlight that the role of clouds as a driver of atmospheric dynamics and climate, well known for giant exoplanets, extends to all sub-stellar bodies.}}



\maketitle

\section{Introduction}\label{sec1}
Brown dwarfs are objects intermediate between exoplanets and stars, with masses ranging from 13 to 80 times the mass of Jupiter. Split into three different spectral types which roughly correlates to effective temperature (L-type (1300 $\leq$T$_{\rm eff}$ $\leq$ 2200 K), T-type (500 $\leq$T$_{\rm eff}$ $\leq$ 1300 K) \cite{kirkpatrick_new_2005} and Y-type (250 $\leq$T$_{\rm eff}$ $\leq$ 500 K) \cite{cushing_discovery_2011}), their atmosphere span a wide range of spectral properties. Especially, on a narrow temperature range around 1300 K, the L/T transition marks a sharp change in observed colours and magnitudes as well as composition with the transition from CO to CH$_{\rm 4}$ as the main C-bearing molecule\cite{vos_astro2020_2019}. During this evolutionary phase, L-type dwarfs appear red whereas T-type dwarfs are bluer, with a brightening observed in the J-band (1.1-1.4 \textmu m) for early T-dwarfs. Distinct explanations for this L/T transition have been proposed, all linked to atmospheric changes. L-dwarfs' thermodynamic conditions favour the presence of silicate and iron condensates high in the atmosphere, which could shape their spectral properties and explain their red colours \cite{tsuji_warm_1999,allard_limiting_2001,ackerman_precipitating_2001,helling_atmospheres_2014,charnay_self-consistent_2018}. The disappearance of these condensates below the photosphere (i.e., the vertical depth probed by observations) as the brown dwarf transitions into a T-type dwarf could explain the observed change in spectral properties. Some 1D models produce a L/T transition following the temperature evolution of brown dwarfs \cite{saumon_evolution_2008,allard_models_2012,charnay_self-consistent_2018}. However, they fail to produce a sharp transition with increasing flux in the J-band. Additional mechanisms related to cloud feedback have been proposed to reinforce the transition, as changes in the cloud cover or vertical mixing induced by cloud convection \cite{marley_patchy_2010,charnay_self-consistent_2018}. Still, such hypotheses require 3D models to be tested. A last possible explanation for the L/T transition is linked to atmospheric convection triggered by chemical instabilities \cite{tremblin_thermo-compositional_2019,tremblin_rotational_2020}. \\ 
Brown dwarfs are rapid rotators (P $\sim$2-20 hours \cite{metchev_weather_2015}), whose atmospheric inhomogeneities, such as patchy cloud cover, are expected to lead to rotational variability in their lightcurves \cite{crossfield_global_2014,metchev_weather_2015}. This variability has been widely observed in multiple brown dwarfs \cite{artigau_photometric_2009,radigan_strong_2014,apai_hst_2013,buenzli_vertical_2012,buenzli_cloud_2015,metchev_weather_2015}, and is probably related to planetary-scale waves \cite{apai_zones_2017}. Specifically, increased variability in the J-band has been observed for L/T transitioning objects \cite{radigan_strong_2014}. Several modelling studies have tackled the variability and L/T transition, mostly focusing on 1-dimensional models \cite{robinson_temperature_2014,morley_water_2014,tan_atmospheric_2019,luna_empirically_2021}. 
However, as atmospheric circulation is a three-dimensional process, understanding the L/T transition and the observed variability requires using 3D Global Climate Models (GCM). Few GCM studies paved the way to such an endeavour. Still, they didn't include clouds, which are thought to play a strong role \cite{showman_atmospheric_2013,zhang_atmospheric_2014,showman_atmospheric_2019,tan_jet_2022,lee_dynamically_2023}. Idealised simulations of early L-dwarfs taking into account cloud formation and their radiative feedback revealed vigorous equatorial dynamics with large-scale zonally propagating waves (Kelvin and Rossby waves), whereas mid-to-high latitudes are dominated by horizontally isotropic vortices \cite{tan_atmospheric_2021-1}. Assuming a given rotation period, the vertical thickness of the cloud layer was found to be greater in the equatorial region than at the poles.  However, this modelling ignored the spectral aspect of radiative transfer and could not thus address the change of colour at the L/T transition. Moreover, their models were tuned to impede convection within the cloud-forming region whereas convective overshoot and mixing by gravity waves can be invoked to maintain dust clouds in the upper levels of hot atmospheres, which disappear when decreasing effective temperature \cite{freytag_role_2010}. Additionally, cloud-convection feedback can drive spontaneous atmospheric variability, as was shown using a Cloud Resolving Model \cite{lefevre_cloud-convection_2022}. To date, an investigation of the drivers of the L/T transition and variability using a GCM  with spectral radiative transfer and self-consistent cloud modelling is lacking. \\ 

\section{Setup}\label{sec2}
To tackle the issue of variability and L/T transition on brown dwarf atmospheres, we use a state-of-the-art 3D multi-wavelength Global Climate Model, the \texttt{generic Planetary Climate Model} (\texttt{generic PCM}) \cite{wordsworth_gliese_2011} coupled with a cloud model which includes condensation, evaporation, transport and radiative effect. This model has recently been adapted for Hot Jupiters \cite{teinturier_radiative_2024} and its adaption to field brown dwarfs is described in Section \ref{sec:methods}. Assuming a rotational period of 5 hours, a gravity of 1000 m.s$^{\rm -2}$, a radius equal to one Jovian radius and silicate clouds made of Mg$_{\rm 2}$SiO$_{\rm 4}$ (dominant type of clouds for a solar metallicity \cite{visscher_atmospheric_2010}), we vary the effective temperature (1500-700 K) and the radius of cloud particles (10-30 \textmu m) to swipe through the L/T transition, and test the effect of the radius of cloud particles on the transition, as particles sizes strongly affect the spectral properties \cite{charnay_self-consistent_2018}. Our simulations are run at a high horizontal resolution (1.4$^{\circ}$ $\times$ 0.94$^{\circ}$ in longitude-latitude), with 80 vertical levels, equally log-spaced in pressure between 1 mbar and 80 bars. This setting is designed to resolve the dynamics, as the characteristic horizontal length scale of dynamical features is close to the deformation radius, which is of the order of 500-5000 km for brown dwarfs \cite{showman_atmospheric_2020}. On the vertical scale, the typical atmospheric scale height is $\sim$ 4 km and needs to be resolved by the model. The runtime is 400 periods, long enough to reach a statistical equilibrium state in the atmosphere.\\
\section{Results}\label{sec3}
\subsection{Atmospheric Structures}

\begin{figure}
    \centering
    \includegraphics[width=\textwidth]{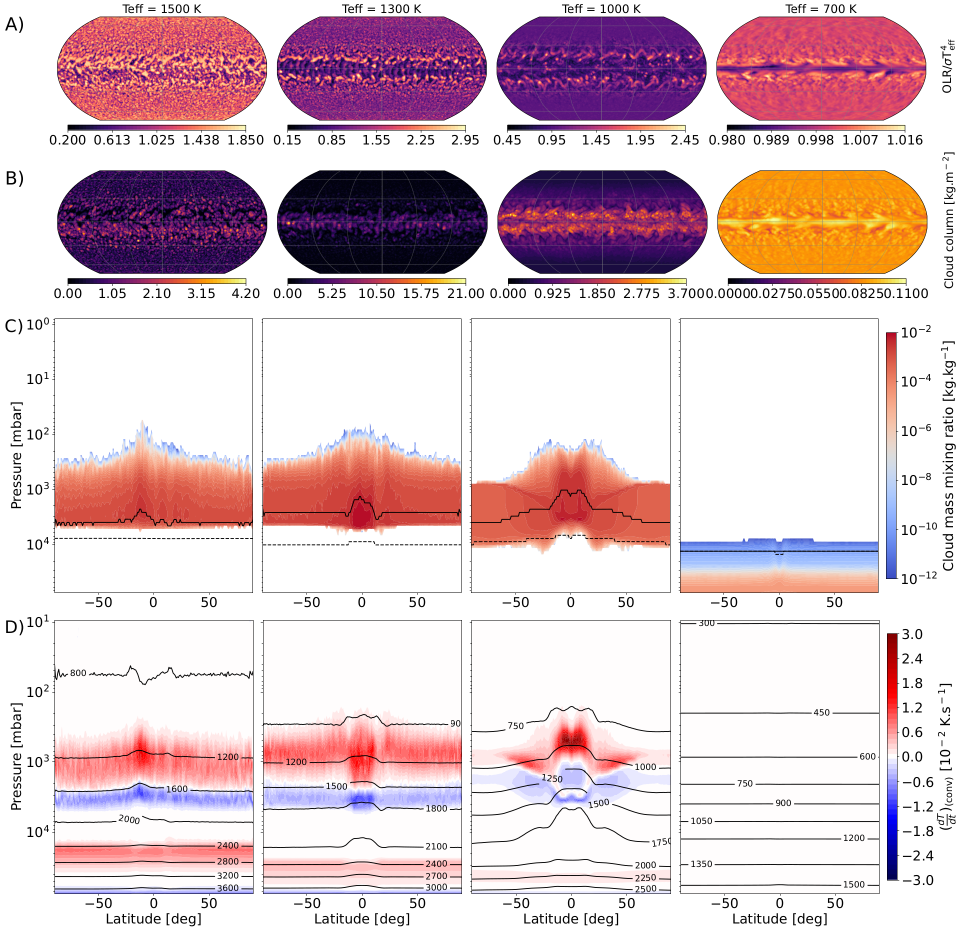}
    \caption{Dynamical states of our simulated brown dwarfs across the L/T transition. (\textbf{A}) Snapshot of the Outgoing Longwave Radiation (OLR) normalised to the mean thermal emission for simulations with cloud particle sizes of 20 \textmu m. (\textbf{B}) Snapshot of the cloud mass columns (vertically integrated). Local variations of cloud columns are correlated with the OLR. We observe patchy cloud covers at all latitudes for T$_{\rm eff} \ge$ 1300 K, a high latitudes depletion of clouds for  T$_{\rm eff}$ = 1000 K and a more uniform cloud deck for T$_{\rm eff}$ = 700 K. (\textbf{C}) Latitude-pressure distribution of the zonal mean of cloud mass mixing ratio. The black solid line denotes the photospheric pressure of the clouds and the atmospheric gas, while the black dashed line is the photospheric pressure of the atmospheric gas only. (\textbf{D})  Latitude-pressure distribution of the convective heating rate (in zonal-mean). Red denotes a positive rate (i.e., warming)  and blue a negative rate (i.e., cooling). The black lines are temperature contours. From left to right, the effective temperatures are 1500, 1300, 1000, and 700 K.}
    \label{fig:1}
\end{figure}

Regardless of the effective temperature of the simulation, two distinct regions emerge in the modelled thermal emission maps (Fig.\ref{fig:1}.A). An equatorial region of strong dynamical activity is present with a latitudinal extent that varies with effective temperature, similarly to \cite{tan_atmospheric_2021, lee_dynamically_2024}. Zonal winds in these regions can reach hundreds of meters per second (see Fig. \ref{fig:ed1}) 
corresponding to the wind speed measured by \cite{allers_measurement_2020}. In this region, the thermal emission is lower than at higher latitudes, due to an increased abundance of clouds. This correlation is seen in Fig.\ref{fig:1}.B which displays the maps of the cloud columns. These maps and the latitude-pressure maps reveal a patchy cover, with more abundant clouds close to the equatorial region than at higher latitudes. \\ 
Indeed, a greater Coriolis parameter leads to weaker vertical velocities and thus a less efficient vertical transport of clouds against gravitational settling \cite{tan_atmospheric_2021}. Thus, thinner clouds populate the higher latitudes. Interestingly, the cloud layer is vertically extended and well-mixed over several scale heights (Fig.\ref{fig:1}.C) leading to an optically thicker layer in the equatorial region. This is due to the radiative effects of clouds, which trigger efficient convective mixing in the cloud layer. Convection is also seen at the lower boundary of the model, as the atmosphere is destabilised by the bottom in our simulations (see bottom panel of Fig.\ref{fig:1}.D). \\ 
At higher latitudes, thermal emission is more horizontally homogenous, as the atmospheric dynamic is dominated by spontaneous vortex formation and eddies over small timescales and lengths. The thermal emission pattern is correlated with cloud patchiness, as cloudy regions emit less due to the increased opacity. The radiative effect of the clouds leads to horizontal gradients of temperature, with a warmer equatorial region and more vertically extended cloud layers (due to the triggering of convection) (Fig. \ref{fig:ed2} and black contours on the bottom panel of Fig.\ref{fig:1}.D). \\
The amplitude of the cloud radiative effects depends on the effective temperature of the simulation, which controls cloud formation. Swiping through the L/T transition (thus lowering effective temperature), we observe a change in the structure of the cloud layer, both vertically and horizontally. From vertically and horizontally extended layers at high effective temperatures (T$_{\rm eff} \ge$ 1300 K) we observe a disruption of the cloud layer at high latitudes and a small vertical sinking for intermediate effective temperatures ($\sim$1000 K) until complete disappearance of the clouds below the observable pressure at cooler effective temperatures ($\sim$ 700 K). For the coolest simulations, no convection is induced by the radiative effect of the Mg$_{\rm 2}$SSiO$_{\rm 4}$ clouds, as is seen in Figure \ref{fig:1}.D. However, other types of clouds could become important at lower temperature \cite{morley_water_2014}.\\
To test whether the clouds have an observable effect, we determined, as a function of latitude, the photospheric pressure at 1.25 \textmu m (in the J-band) as the level having a temperature equal to the brightness temperature at this wavelength. This level corresponds to the region probed by the observed thermal emission.  
To distinguish from a case where no clouds affect the photospheric levels, we pursued the simulations for one additional timestep where we turned off the radiative effect of clouds and only included the gas opacity. This is shown as black solid lines (when the clouds are radiatively active) and black dashed lines (when the clouds are transparent) in Fig.\ref{fig:1}.C. For an effective temperature above 1000 K and cloud particle sizes of 20 \textmu m, the photospheric pressure is set by the contribution of the cloud opacity. It is thus at lower pressure (higher altitudes). In contrast, for lower effective temperatures, cloud opacity does not contribute to the location of the photospheric pressure (which is thus located deeper in the atmosphere). This is due to clouds sinking below the cloudless photospheric pressure in these cases, and their presence has virtually no observable impact. \\ 

\subsection{Link with observed magnitudes and colours}

\begin{figure}
    \centering
    \includegraphics[width=\textwidth]{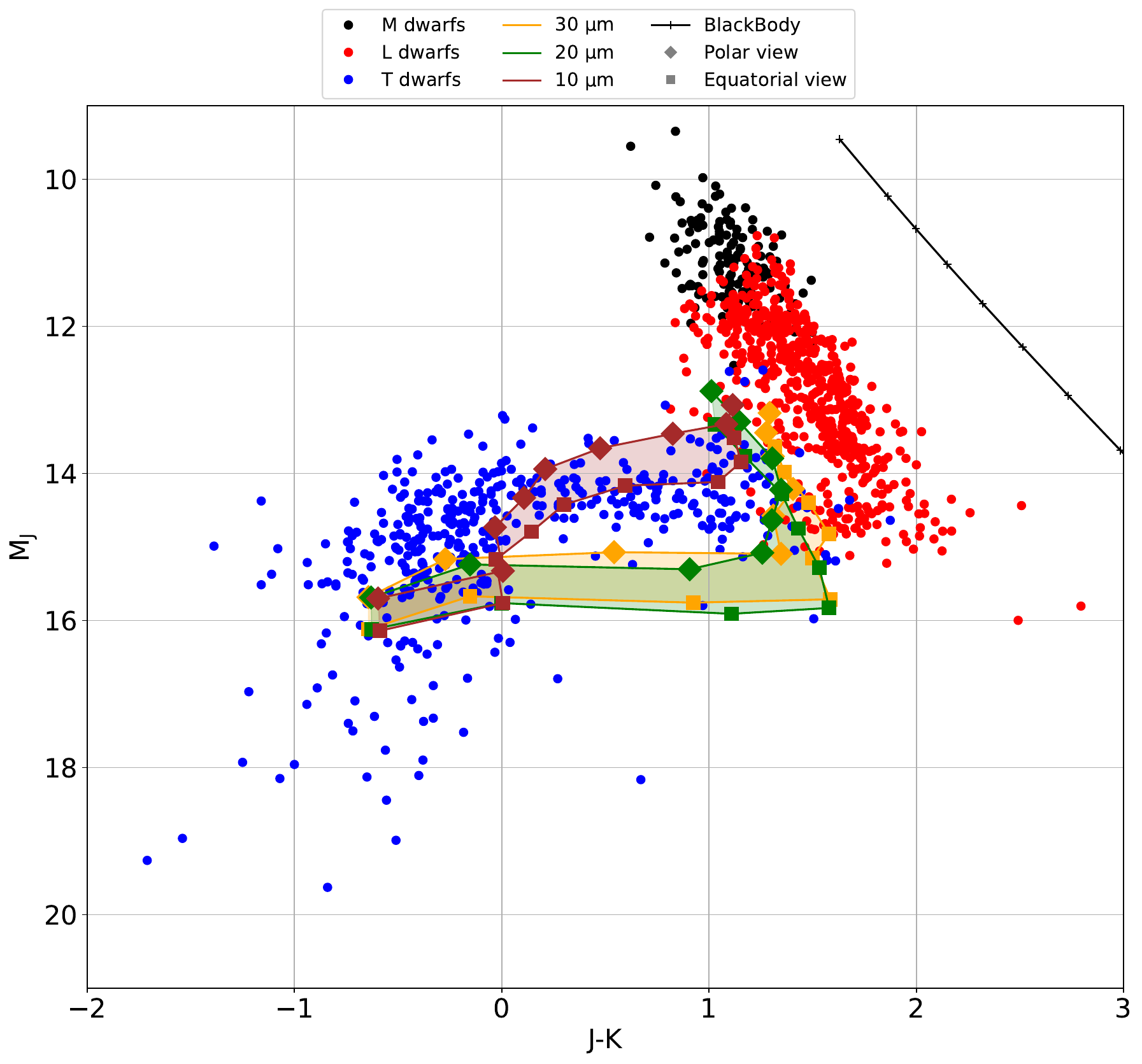}
    \caption{Colour-Magnitude diagram with J-K colours plotted against absolute J magnitude in the MKO system. Background data include M-dwarfs (black points), L dwarfs (red points) and T dwarfs (blue points) taken from \cite{best_ultracoolsheet_2024}. Overlaid are our simulations with different cloud particle sizes (10 \textmu m in brown, 20 \textmu m in green, and 30 \textmu m in yellow). Each point represents an effective temperature, ranging from 1500 to 700 K with steps of 100 K. The shaded area corresponds to the range resulting from the variation of the viewing angle from an equatorial view (dotted symbols) to a polar view (diamond symbol). The black line with crosses represents a blackbody with a temperature ranging from 2200 K to 700 K (outside the range of the figure).}
    \label{fig:2}
\end{figure}

This sinking of the cloud layer below the observable pressure for the cooler simulations results in the observation of clear-sky objects, as the opaque cloud layer does not obscure the photospheric pressure. This leads to a transition in the observed colours, as shown in Figure \ref{fig:2}. The colour-magnitude diagram (CMD) displays the colours (in J-K) and absolute J-magnitude from our simulations computed by varying the viewing angle of the brown dwarf, from an equatorial to a polar view. In all cases, the equatorial views appear fainter and redder, as a consequence of the additional cloud opacity, in agreement with observations \cite{vos_viewing_2017}. In our GCM simulations, atmospheric variability (and resulting colour changes) with effective temperature are found to be driven by the radiative effect of clouds and naturally lead to a shape of the L/T transition compliant with observations: This strongly suggests that clouds play the main role in shaping the L/T transition.

Apart from the blueing, a slight brightening for T-dwarfs is an outcome of our simulations for cloud particle sizes bigger than 10 \textmu m. This sharp transition is due to the greenhouse effect and the efficient mixing (due to convection) of clouds. The size of the cloud particle only mildly impacts the shape of the simulated transition except for the 10 \textmu m case, where the effective temperature of transition exactly matches the observations. This suggests that fine-tuning the physical parameters of the simulations should yield quantitative results that match the absolute level of the observations as gravity, rotation period, atmospheric metallicity, cloud particle sizes, additional condensates, etc., also contribute to the observed colours and magnitude. \\
The mechanism highlighted here is an intrinsically 3D effect, that cannot be captured by 1D models. Instead of arbitrary patchy cloud cover \cite{marley_patchy_2010}, our simulations show that it is the combination of the cloud radiative effects and inhomogeneous cloud cover between the equatorial and high latitudes (due to temperature gradients between these regions) that drives the L/T transition. With decreasing effective temperature, the horizontal and vertical thicknesses of the cloud cover decrease, leading to a blueing of the objects.

\subsection{Variability}

\begin{figure}
    \centering
    \includegraphics[width=\textwidth]{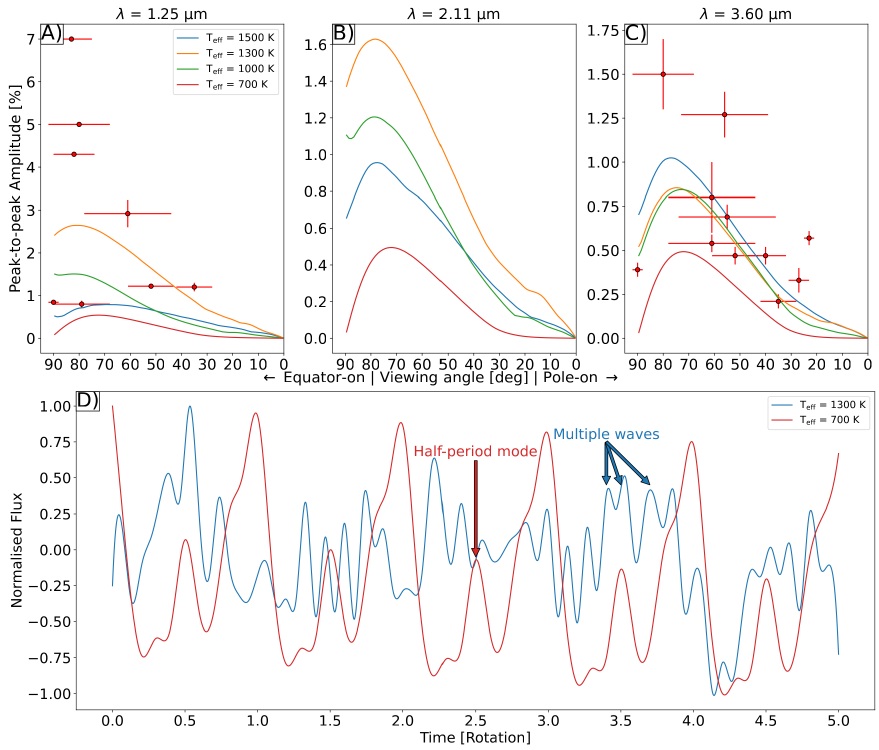}
    \caption{ Time, spectral and spatial variability across the L/T transition. The top row (\textbf{A} to \textbf{C}) displays the peak-to-peak amplitude as a function of the viewing angle, for three spectral bands (J-band (\textbf{A}), K-band (\textbf{B}) and Spitzer IRAC Channel 1 (\textbf{C}) respectively). Red dots are variability measurements taken from \cite{vos_viewing_2017,vos_variability_2018} (uncertainties can be found in those papers) and colours represent different effective temperatures, ranging from 1500 to 700 K.  (\textbf{D}) shows the normalised white lightcurves as a function of time for models with effective temperature of 1300 and 700 K. Temporal variability is seen in the warmer model, due to enhanced cloud and wave activity. The brown dwarf is viewed equator-on in this panel. Cloud particle sizes are 20 \textmu m. }
    \label{fig:3}
\end{figure}

From our thermal emission maps of Fig.\ref{fig:1}.A and our CMD of Fig.\ref{fig:2},  we observe that spatial variability is a natural characteristic of brown dwarf atmospheres. As shown with the range of the thermal emission maps, this spatial variability is enhanced around the L/T transition and is weaker for late T-dwarfs, as a consequence of cloud radiative effect (or lack thereof). Spectral and time variability are also an outcome of our modelling, as shown in Fig.\ref{fig:3}. Variability is enhanced at the L/T transition \cite{radigan_strong_2014} and is maximal at near-infrared wavelengths \cite{metchev_weather_2015}, whereas mid-infrared spectral windows are less sensitive to cloud-induced variability. Moreover, the viewing angle is key for variability detection, as equator-on objects are dramatically more variable than polar-on objects as has been observed by \cite{vos_viewing_2017,suarez_ultracool_2023} and our simulations reproduce this. In particular, comparing our simulations with observed variability measurements yields a good match in the Spitzer Channel 1 bands (Fig.\ref{fig:3}.C) but is underestimated in the J-band (Fig.3\ref{fig:3}A). However, the measurements from \cite{vos_viewing_2017,vos_variability_2018} encompass various effective temperatures, metallicity, radius, and gravity, translating to a dispersion of variability not seen in our simulations which only explore a subset of the parameter space. \\

\begin{figure}
    \centering
    \includegraphics[width=\textwidth]{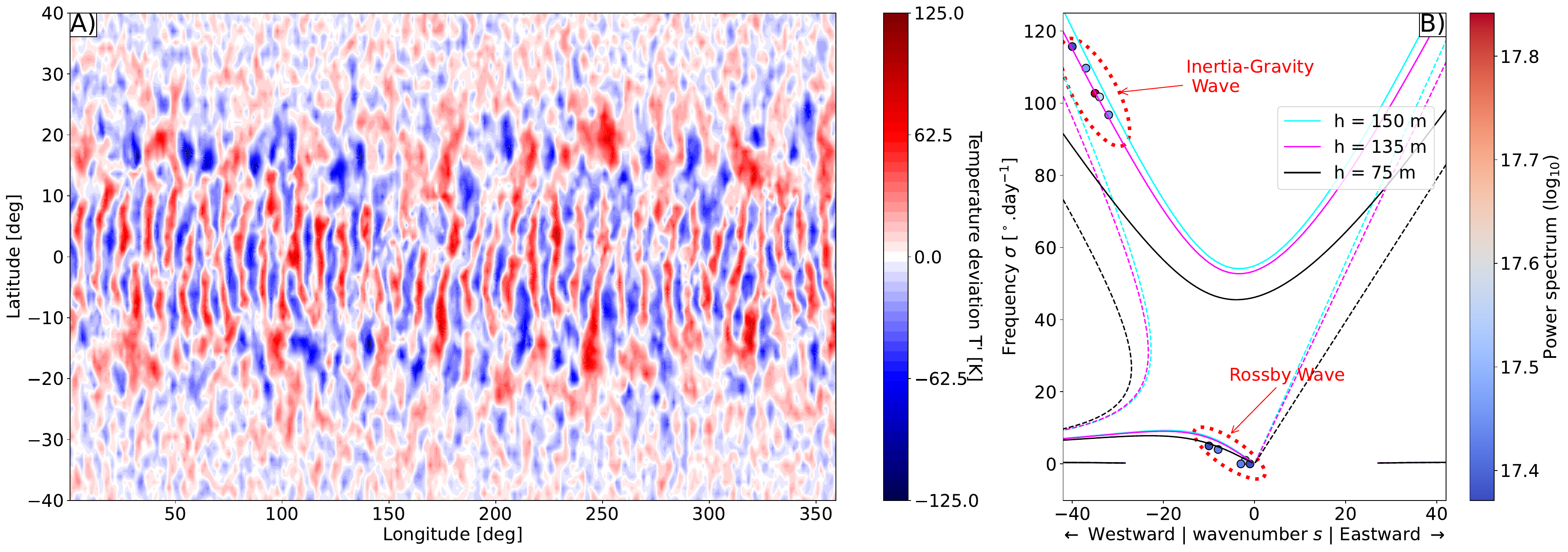}
    \caption{Equatorial waves diagnostic for an effective temperature of 1300 K. (\textbf{A}) is a map of the temperature deviations T' from the zonal mean at a pressure of 1.5 bars. Rossby wave patterns modulated by higher frequency waves can be seen in the equatorial region. Equatorially symmetric behaviour between the two hemispheres is prominent. (\textbf{B}) is the symmetric component of our wave spectral analysis, close to the equator. Solid and dashed lines represent theoretical dispersion relationships for Rossby and Kelvin waves, and data points are the ten dominant modes calculated following the methodology described in Section 5.5. Different colours represent different equivalent depths. Inertia-Gravity and Rossby waves are detected with equivalent depths of $\sim$ 100~m.}
    \label{fig:4}
\end{figure}

Looking at Fig.\ref{fig:3}.D, which displays the simulated normalised (between -1 and 1) lightcurves as a function of time, we observe a regular double-peaked shape for the lowest temperature, whereas the behaviour is more chaotic and variable in time for warmer temperatures. In particular, a half-period mode is present at low effective temperature, similar to what has been previously observed with long-term monitoring of a few brown dwarfs \cite{apai_zones_2017,fuda_latitude-dependent_2024}. This mode is explained by one global-scale zonally propagating Rossby wave in each hemisphere, which mimics a zonal wavenumber  $k=2$ wave as established by previous GCM simulations \cite{tan_jet_2022}. For warmer simulations, multiple waves disrupt the sinusoidal shape of the lightcurve, with periods both smaller and longer than the rotation period. The triggering of these waves is easily understandable as warmer simulations feature equatorial convective clouds, which may generate Rossby waves, as convective heating induces fluctuations in the large-scale horizontal divergence. These represent a source of vorticity fluctuations, leading to Rossby waves \cite{sardeshmukh_generation_1988,schneider_formation_2009}. Rossby waves indeed occur in our GCM simulations, as can be seen for an effective temperature of 1300 K in Fig.\ref{fig:4}.A, exhibiting a pattern reminiscent of simulations of Jupiter and Saturn \cite{liu_mechanisms_2010, spiga_global_2020}; as in these Jupiter/Saturn simulations, 
our 1300~K simulations exhibit equatorial super-rotation often associated with equatorial Rossby waves. Using spectral analysis and a decomposition onto a symmetric/antisymmetric component \cite{wheeler_convectively_1999}, we highlight the existence of both low-frequency equatorial Rossby waves and high-frequency inertia-gravity waves (Fig.\ref{fig:4}.B). This enhanced wave activity linked to increased cloud radiative effects might be responsible for the observed transition from low to high amplitude lightcurves and transition from single to multiple-peaked lightcurves in a few rotations \cite{artigau_photometric_2009,yang_extrasolar_2016,artigau_variability_2018}.
\section{Conclusion}\label{sec4}
We have shown that the radiative effect of clouds in the atmosphere of brown dwarfs can reproduce a sharp L/T transition and explain the observed spectral, spatial and temporal variability of these objects. Our simulations point to an enhanced dynamical and wave activity in the equatorial region, due to increased cloudiness. The radiative forcing by clouds induces convective mixing that allows for a well-mixed and optically thick cloud layer, and maintains the dynamical activity. Away from the equatorial region, cloud formation and disruption are stochastic in time and space, and the lack of organised structure leads to weaker variability, as postulated originally by \cite{ackerman_precipitating_2001}. The viewing angle is a key parameter in the detection of variability, as equator-on brown dwarfs are more variable than their pole-on counterparts, and also appear redder and fainter. Thus, cloud radiative effect is demonstrated to be a plausible driver of the L/T transition and spectral variability in brown dwarfs.
The James Webb Space Telescope and Extremely Large Telescopes will help to further characterize cloud dynamics on brown dwarfs, using variability monitoring and atmospheric mapping\cite{biller_clouds_2023,plummer_mapping_2023}.
Finally, this study highlights the importance of cloud feedback on the dynamics and evolution of planetary atmospheres. The breakup of clouds for low effective temperatures in brown dwarfs, as the cloudy photosphere dives below the gas photosphere shows some analogy with the breakup of stratocumulus decks in the presence of high greenhouse gases on Earth, as predicted by some global and regional models for the Earth's past and future climates \cite{schneider_possible_2019, goldblatt_earths_2021, leconte_increased_2013, yan_cloud_2022}. These similar cloud behaviours from climate models under different conditions may suggest a general reduction of the cloud cover with increasing greenhouse gas concentration.
\newpage
\section{Methods}\label{sec:methods}
\subsection{Atmospheric Model}
In this work, we use the generic Planetary Climate Model (\texttt{generic PCM}) \cite{wordsworth_gliese_2011,charnay_3d_2015,charnay_3d_2015-1,spiga_global_2020,bardet_joint_2022,turbet_water_2023,milcareck_radiative-convective_2024,teinturier_radiative_2024,boissinot_global_2024} on which we made several modifications. This model couples two different modules. The first one is a dynamical core and the second is a physical package. The dynamical core used in this study is DYNAMICO, a core tailored for massively parallel computations \cite{dubos_dynamico-10_2015}. This core solves the primitive equations of meteorology assuming a shallow atmosphere. The sphere is discretised as a quasi-uniform icosahedral C-grid. A three-dimensional Hamiltonian formalism is used, which enforces energy conservation. Subgrid-scale dissipation in the horizontal dimension is also included as an additional term in the vorticity, divergence and thermodynamic equations to prevent the accumulation at scales close to the grid resolution. For all the simulations presented here, the horizontal resolution is 1.40625 $\times$ 0.9375$^{\circ}$ in longitude-latitude (equivalent to 256 $\times$ 192 for a regular longitude-latitude grid). Time-integration is performed using an explicit Runge-Kutta scheme, chosen for stability and accuracy. Vertical coordinates are sigma coordinates, defined as the ratio of the pressure to the pressure at the bottom of the model (set at 80 bars). We use 80 levels from the bottom of the model to the top of the atmosphere, which is set at a pressure of 1 mbar. Extension to lower top pressure was found to be difficult due to the appearance of numerical instabilities. \\
The inputs and outputs of DYNAMICO are handled with the XIOS library (XML Input/Output Server)\cite{meurdesoif_xios_2020}. This library manages all input/output operations independently of the numerical integrations, which allows for great numerical efficiency when the computation is distributed over tens to hundreds of cores.  During runtime, XIOS converts all the dynamical fields computed on the quasi-uniform icosahedral C-grid onto a regular longitude-latitude grid, using finite-volume weighting functions computed at the beginning of the run. \\ 
The dynamical core is coupled to the physical package tailored for hot gaseous atmospheres and presented in \cite{teinturier_radiative_2024}. This package includes a radiative transfer scheme based on the k-correlated method for molecular opacities\cite{goody_correlated-k_1989} and the two-stream equations, including collision-induced absorption of H$_{\rm 2}$-H$_{\rm 2}$ and H$_{\rm 2}$-He. The implementation is based on the algorithm of \cite{toon_rapid_1989} which incorporates a shortwave and a longwave channel to solve the radiative transfer equation. The shortwave channel is traditionnaly used for the stellar flux (which doesn't exist in our case) and the longwave channel for the planetary emission. For the particular case of field brown dwarfs, we deactivate the shortwave channel, by setting the number of bins in this channel to one and setting the according k-coefficients to zero. Traditionally, the Rayleigh scattering by H$_{\rm 2}$ and He is only included in the shortwave channel and we modified the code to include it in the longwave channel, as the shortwave channel is not used. Moreover, to ensure that no stellar flux whatsoever reaches the brown dwarf atmosphere, we set the stellar flux at 1 AU to be equal to 0 W.m$^{\rm -2}$. This is mandatory as the \texttt{generic PCM} has a built-in host star, which is not relevant here. The model also includes a convective adjustment scheme, which relaxes the temperature profile toward the adiabatic lapse rate if an unstable lapse rate is encountered in a vertical column \cite{hourdin_meteorological_1993} and a vertical diffusion scheme for small-scale turbulence following \cite{mellor_development_1982}.  At the bottom of the model, a latitude-independent heat flux is applied. \\ 
The cloud scheme is identical to the one presented in \cite{teinturier_radiative_2024} and summarised here. Cloud and condensable vapour are represented by tracers that can be advected with the hydrodynamical flow using a 'Van-leer I' finite volume scheme, which is built into the dynamical core \cite{hourdin_use_1999}. In each layer and grid cell, the saturation pressure of a given species is computed using the Clausius-Clapeyron law. For saturated layers and grid cells, the whole amount of condensable vapour available is condensed into a cloud. Otherwise, an equilibrium is computed between solid and vapour phases following: \\
\begin{equation}
    q_{sat} = \epsilon \frac{P_{sat}}{P-(1-\epsilon)P_{sat}} 
\end{equation}\\
where $\epsilon_i = m/\mu$ with $m$ the molecular weight of the condensable species, $\mu$ the mean molecular atmospheric weight of the background gases, $q_{\rm sat}$ this specific concentration at saturation of the condensable species, $P$ the pressure and $P_{\rm sat}$ the saturation vapour pressure. In this work, we only include clouds made of Mg$_{\rm 2}$SiO$_{\rm 4}$ and the saturation vapour pressure law used is taken from Table 2 of \cite{visscher_atmospheric_2010}. Latent heat release from cloud formation is taken into account. \\
Clouds can experience sedimentation at a terminal velocity and coagulation or coalescence is neglected. The terminal velocity is computed using the non-linear corrected Stokes law of \cite{fuchs_mechanics_1965,ackerman_precipitating_2001}: 
\begin{equation}
    V_f = \frac{2\beta a^2g(\rho_{\rm Mg_{\rm 2}SiO_{\rm 4}}-\rho)}{9 \eta}
\end{equation}
where $a$ is the particle radius, $g$ is the gravity, $\rho$ is the atmospheric density, $\rho_{\rm Mg_{\rm 2}SiO_{\rm 4}}$ is the density of Mg$_{\rm 2}$SiO$_{\rm 4}$ and is taken equal to 3270 kg.m$^{\rm -3}$, $\eta$ is the atmospheric viscosity and $\beta$ the non-dimensional Cunningham slip factor: 
\begin{equation}
    \beta = 1 + K_n(1.256+0.4\exp{(-1.1/K_n)})
\end{equation}
where $K_n$ is the Knudsen number, given by $K_n = \lambda/a$ with
\begin{equation}
    \lambda = \frac{k_{\rm B}T}{\sqrt{2}\pi d^2P}
\end{equation}
where $d$ is the effective gas molecular radius.

Viscosity is computed as a weighted expression of the viscosity of H$_{\rm 2}$, He, H$_{\rm 2}$O following the relationships derived by \cite{rosner_transport_1986,petersen_properties_1970,sengers_representative_1984}: 
\begin{equation}
    \eta = q_{\rm H_{\rm 2}}\Big(2 \times 10^{-7} T^{0.66}\Big)+ q_{\rm H_{\rm e}}\Big(1.9 \times 10^{-5}(T/273.15)^{0.7} \Big) + 8 \times 10^{-6} q_{\rm H_{\rm 2}O}
\end{equation}
The radiative effects of clouds are also taken into account. As we use the plane-parallel two-stream framework for the radiative transfer, cloud particle radiative properties are fully described by the extinction efficiency Q$_{e}$, the single scattering albedo $\varpi$ and the asymmetry parameter g$_0$. These three quantities were computed offline as a function of size and wavelength and used in \cite{charnay_self-consistent_2018,teinturier_radiative_2024}. During the simulations, these parameters are interpolated at each time-step and for each grid cell, based on a prescribed mean particle radius $\overline{a}$ and assuming a log-normal size distribution as is common for these atmospheres \cite{tan_atmospheric_2021-1,teinturier_radiative_2024,lee_dynamically_2024}. For each atmospheric layer, the optical depth is computed as : 
\begin{equation}
    d\tau = \frac{3 Q_e}{4 \rho_{\rm Mg_{\rm 2}SiO_{\rm 4}}\overline{a}}q_{\rm Mg_{\rm 2}SiO{\rm 4}}\frac{dp}{g}
\end{equation}
where $dp$ is the pressure thickness of the layer, $g$ the constant-with-altitude acceleration of gravity and $q_{\rm Mg_{\rm 2}SiO{\rm 4}}$ the specific concentration of Mg$_{\rm 2}$SiO$_{\rm 4}$ cloud in the layer. We fixed the width of the distribution at $\sqrt{0.1}$. \\

\subsection{Model Initialisation}
To compute the k-correlated tables used by the 3D model, we start from the k-coefficients computed in \cite{blain_1d_2021} and run the 1D radiative-convective code \texttt{Exo-REM} \cite{charnay_self-consistent_2018} with disequilibrium chemistry, using a parameterization for the eddy diffusion coefficient derived by \cite{ackerman_precipitating_2001}. Solar elemental abundance of 13 species is assumed (H$_{\rm 2}$O, CO, CH$_4$, CO$_2$, FeH, HCN, H$_2$S, TiO, VO, Na, K, PH$_3$ and NH$_3$), which sets the chemistry for the 3D simulations (see Fig. \ref{fig:sup1}). We create mixed k-tables depending on pressure and temperature using the \texttt{exo\_k} package \cite{leconte_spectral_2021} from the calculated thermal profile and vertical abundance profiles of each species. These mixed k-tables are computed using 16 Gauss-Legendre quadrature points and 15 spectral bins between 0.261 and 325 \textmu m. The 1D version of the \texttt{generic PCM}, which consists of the physical package only on one vertical column, is then run using these mixed k-tables. The 3D shell is initialised with the output thermal profile on each grid point. Using this thermal profile, condensable vapour is horizontally uniformly added to the shell using the analytical profiles for Mg$_{\rm 2}$SiO$_{\rm 4}$ derived by \cite{visscher_atmospheric_2010}. One issue with simulations of field brown dwarfs is the absence of inhomogeneities in external forcing, as no radiation heats the atmosphere from above and the internal heat flux is horizontally homogenous. As noted by \cite{tan_jet_2022}, a random perturbation is needed to drive the dynamics; otherwise, only radiative-convective equilibrium will be observed. Thus, we impose at the beginning of each simulation a random perturbation of up to 10 K on the thermal profile to create horizontal inhomogeneities. We insist on the fact that this perturbation is only applied at the beginning of the simulations and that the radiative effect of the clouds will then drive the dynamics. The whole  procedure described above is repeated for each effective temperature used in this study, from 1500 to 700 K. \\

\subsection{Parameters of the simulations}
We use three different radii of particles for the Mg$_{\rm 2}$SiO$_{\rm 4}$ clouds, 10, 20 and 30 \textmu m, and we vary the internal temperature (and thus the internal heat flux) from 700 K to 1500 K with a step of 100 K. For all the simulations, we use a fixed radius of 1 Jovian radius, a rotational period of 5 hours and gravity of $\log{g}$ = 5 (in cgs units). These parameters were chosen as representative of the brown dwarf population. The choice of radii for the cloud particles was motivated by a previous study \cite{cooper_modeling_2003}. \\
Each simulation was run for 400 rotations, allowing it to reach a statistical equilibrium. The nominal hydro-dynamical time-step is 20 seconds. The radiative/physical time-step differs depending on the effective temperature of the simulations. For T$_{\rm eff} \leq$1100 K, the time-step is 60 seconds and is 40 seconds otherwise. Two simulations (T$_{\rm eff}$ = 1400, 1300 K for 10 \textmu m cloud particles) were run with shorter time-steps to avoid numerical instabilities. This is due to an increase in cloud radiative effect due to thicker clouds. Then, the dynamical time-step used is 10 seconds and the physical/radiative time-step is respectively 20 and 30 seconds for T$_{\rm eff}$ = 1300 and 1400 K.\\
To mitigate the energy cascade toward the small, unresolved horizontal scales, we use a hyperdiffusion time constant of 10000 seconds, which translates to a weak term on the vorticity, divergence,e, and thermodynamic equations. This choice was motivated by past modelling experiences with DYNAMICO \cite{spiga_global_2020,bardet_global_2021}. \\
We ran the simulations on 96 Intel Platonium 8168 processors. Each simulation takes a few weeks to complete, depending on the effective temperature and the cloud particle size. We roughly estimate that our computing resources used around 6.7 teqCO$_{\rm 2}$ when taking into account developing, testing and scientific simulations.\\ 

\subsection{Magnitude and colour computation}
To compute the colour-magnitude diagram, we post-process all our simulations with a higher spectral resolution. Similarly to what was done to generate the k-tables at low spectral resolution, we use the \texttt{exo\_k} package and compute 1069 spectral bins between 0.261 and 325 \textmu m using the same opacity sources as for the low-resolution k-tables. We then run one extra time-step of the 3D model at this high spectral resolution. \\
We compute the disk-averaged emission spectra depending on the viewing angle (which we vary from the equator to poles) of the brown dwarfs using the visibility kernel of \cite{cowan_light_2013}. Then, we integrate the disk-averaged spectra and weigh them by the transmission function of the J-band (1.14-1.36 \textmu m) or K-band (1.96-2.41 \textmu m) MKO filter. \\
Finally, as brown dwarfs cool over time and contract in size, we use the evolutionary law of \cite{burrows_theory_2001} for the radius, which scales as T$_{\rm eff}^{0.11}$, following: 
\begin{equation}
    R \sim 6.7 \times 10^4 \rm{km} \Bigg(\frac{10^5}{g}\Bigg)^{0.18}\Bigg(\frac{T_{eff}}{1000\rm{ K}}\Bigg)^{0.11}
\end{equation}
where $g$ is the surface gravity in cgs units, $R$ the radius of the brown dwarf and $T_{\rm eff}$ the effective temperature in K.

\subsection{Spectral analysis of planetary waves}
Strong wave activity is seen in the equatorial region of our simulations, and their characterisation is needed to understand their forcing on the circulation. We use the Wheeler and Kiladis method \cite{wheeler_convectively_1999} which is extensively used to study the stratospheric equatorial waves on Earth \cite{kiladis_convectively_2009,maury_presence_2014}. This technique has also been applied to GCM simulations of Saturn to diagnose the equatorial stratospheric and tropospheric waves \cite{spiga_global_2020,bardet_global_2021,bardet_joint_2022}. This method consists of a two-dimensional Fourier transform from the longitude-time ($\lambda$-$t$) space to the zonal wavenumber-frequency ($s$-$\sigma$) space of the symmetric and antisymmetric component of a geophysical field $X$ about the equator. For a given latitude $\phi$, these two components are defined as: 
\begin{equation}
    X_S = \frac{1}{2}\Big(X_{\phi N}+X_{\phi S}\Big) 
\end{equation}
\begin{equation}
     X_A = \frac{1}{2}\Big(X_{\phi N}-X_{\phi S}\Big) 
\end{equation}
In our case, we use the OLR field. Our technique is similar to the one used by \cite{tan_atmospheric_2021-1}. The symmetric component is used to determine planetary-scale Rossby and Kelvin waves. In contrast, the antisymmetric component is used to determine planetary-scale mixed Rossby gravity and eastward inertia-gravity waves. \\ 
To help identify the different waves in the equatorial regions of our simulations, theoretical dispersion relations are superimposed on the power spectra. These relations are based on the linear waves theory. For equatorial waves, the relations are \cite{maury_presence_2014}:
\begin{equation}
    \gamma \sigma^2-s^2-s/\sigma = \sqrt{\gamma}\big(2\nu+1\big)
\end{equation}
with $\nu$ the meridional mode number, $\gamma$ the Lamb parameter defined as 
\begin{equation}
    \gamma = \frac{4a^2\Omega^2}{gh}
\end{equation}
where $a$ is the radius of the brown dwarf, $\Omega$ the rotation rate, $g$ the surface gravity and $h$ is an equivalent depth associated with the vertical wavenumber $m$: 
\begin{equation}
    m^2 = \frac{N^2}{gh}-\frac{1}{4H^2}
\end{equation} 
where $N$ is the Brunt-V{\"a}is{\"a}l{\"a} frequency and $H$ the scale height. These equations define the type of waves via the value of $v$. For example, Rossby waves are defined by $v$ = 1, Kelvin waves for $v$=-1, mixed Rossby gravity waves for $v$=0 and eastward inertia gravity waves for $v$ =2. The equivalent depth $h$ is linked to the vertical wavenumber $m$ of the waves following: 
\begin{equation}
    m^2 = \frac{N^2}{gh}-\frac{1}{4H^2}
\end{equation}
\\
Our spectral analysis code has been extensively tested on Saturn \cite{spiga_global_2020,bardet_global_2021,bardet_joint_2022} and on well-defined diurnal tides and Kelvin waves simulated in the Martian atmosphere. As our wind speeds are strong, we correct for the Doppler effect due to the background wind.

\section*{Data Availability}
The data used in this study are available on this zenodo repository: \cite{teinturier_clouds_2025}

\section*{Code Availability}
The \texttt{generic PCM} is available here: \url{https://trac.lmd.jussieu.fr/Planeto} and the \texttt{DYNAMICO} core is available here: \url{http://forge.ipsl.jussieu.fr/dynamico}.

\section*{Acknowledgements}
L.T. thanks X. Tan for instructive discussions about the initialisation of field brown dwarf simulations during the ExoClimes IV conference and for his encouragement in pursuing this work. L.T thanks M. Lefèvre for pertinent discussions on cloud convection. L.T. strongly thanks the MesoPSL engineers at Observatoire de Paris for their great support, reactivity and help in troubleshooting the installation of the DYNAMICO core on the machines. The authors thank the three anonymous referees, whose comments and questions led to a great improvement and clarity of the manuscript. \\

\section*{Author Contributions Statement}
L.T. led the project, developed the model, ran the simulations and drafted the manuscript. B.C. and A.S. contributed to the project supervision, the model development, the physical interpretation and the writing of the manuscript. B.B. contributed to the physical interpretation of the simulations and the writing of the manuscript.
\section*{Competing Interests Statement}
The authors declare no competing interests.

\section*{Figures}

\renewcommand{\thefigure}{Extended Data \arabic{figure}}
\setcounter{figure}{0}
\begin{figure}
    \centering
    \includegraphics[width=\textwidth]{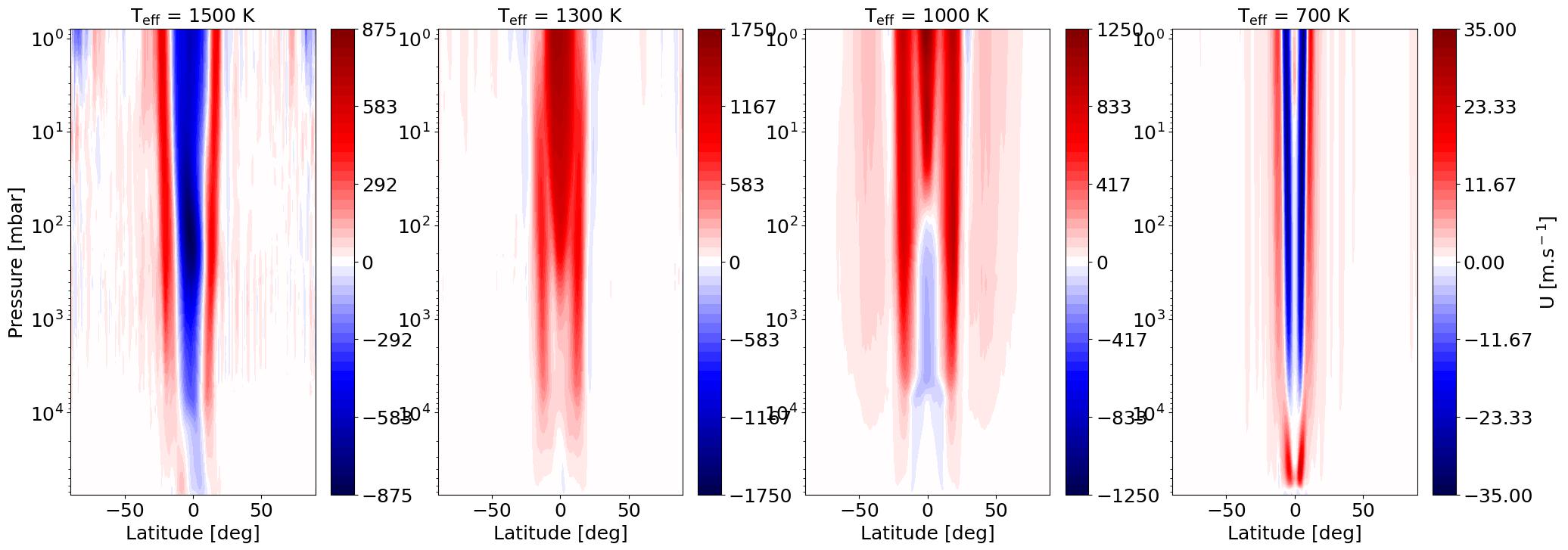}
    \caption{Snapshot of the latitude-pressure distribution of the zonal mean zonal wind. Positive values (in red) indicate an eastward propagation, whereas negative values (in blue) indicate a westward propagation. Decreasing effective temperature leads to weaker winds, because of weaker thermal feedback of the clouds.
}
    \label{fig:ed1}
\end{figure}
\begin{figure}
    \centering
    \includegraphics[width=\textwidth]{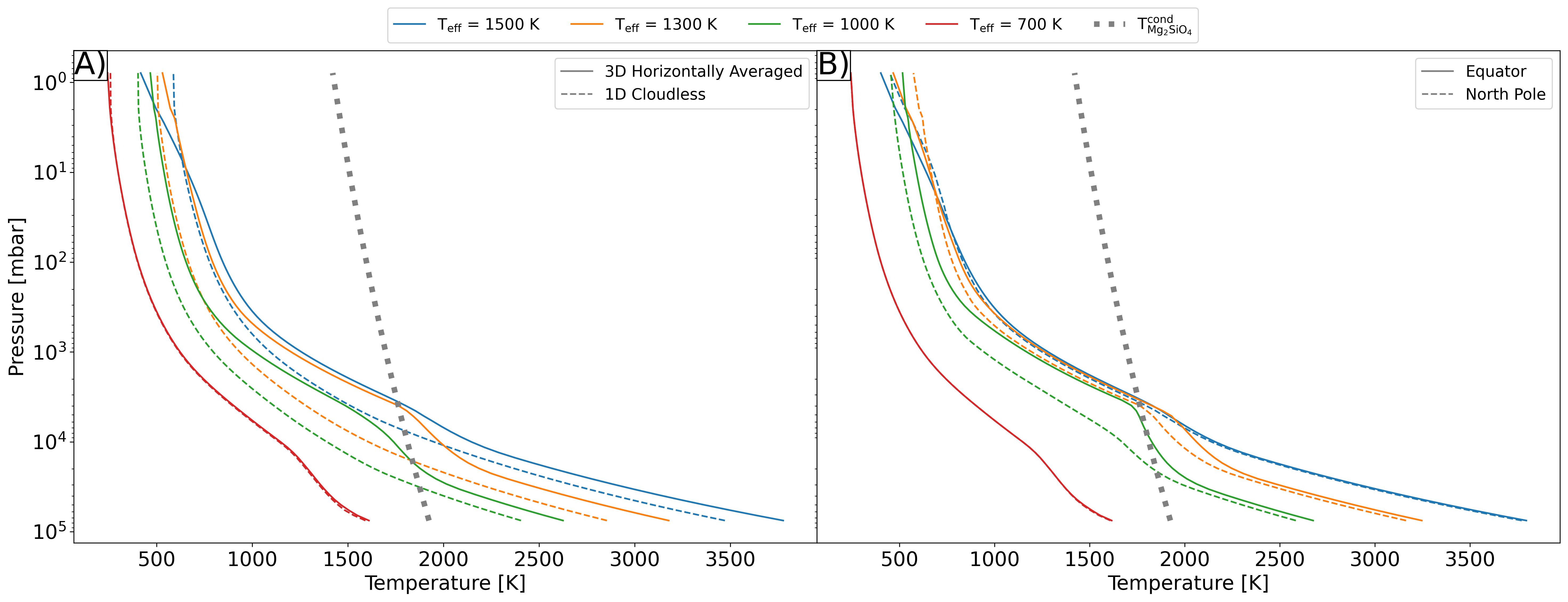}
    \caption{\textbf{(A)} Time-averaged and horizontally averaged temperature profiles from the cloudy 3D simulations (solid lines), and 1D cloudless temperature profiles (dashed lines). \textbf{(B)} Time-averaged and zonal mean temperature profiles at the equator (solid lines) and at the North pole (dashed lines). For both panels, each colour represents a different effective temperature and the grey, thick dotted line shows the condensation temperature for silicate (Mg$_{\rm 2}$SiO$_{\rm 4}$) clouds. Where the temperature profile crosses the condensation temperature, clouds form or evaporate. 
}
    \label{fig:ed2}
\end{figure}
\renewcommand{\thefigure}{Supplementary \arabic{figure}}
\setcounter{figure}{0}
\begin{figure}
    \centering
    \includegraphics[width=\textwidth]{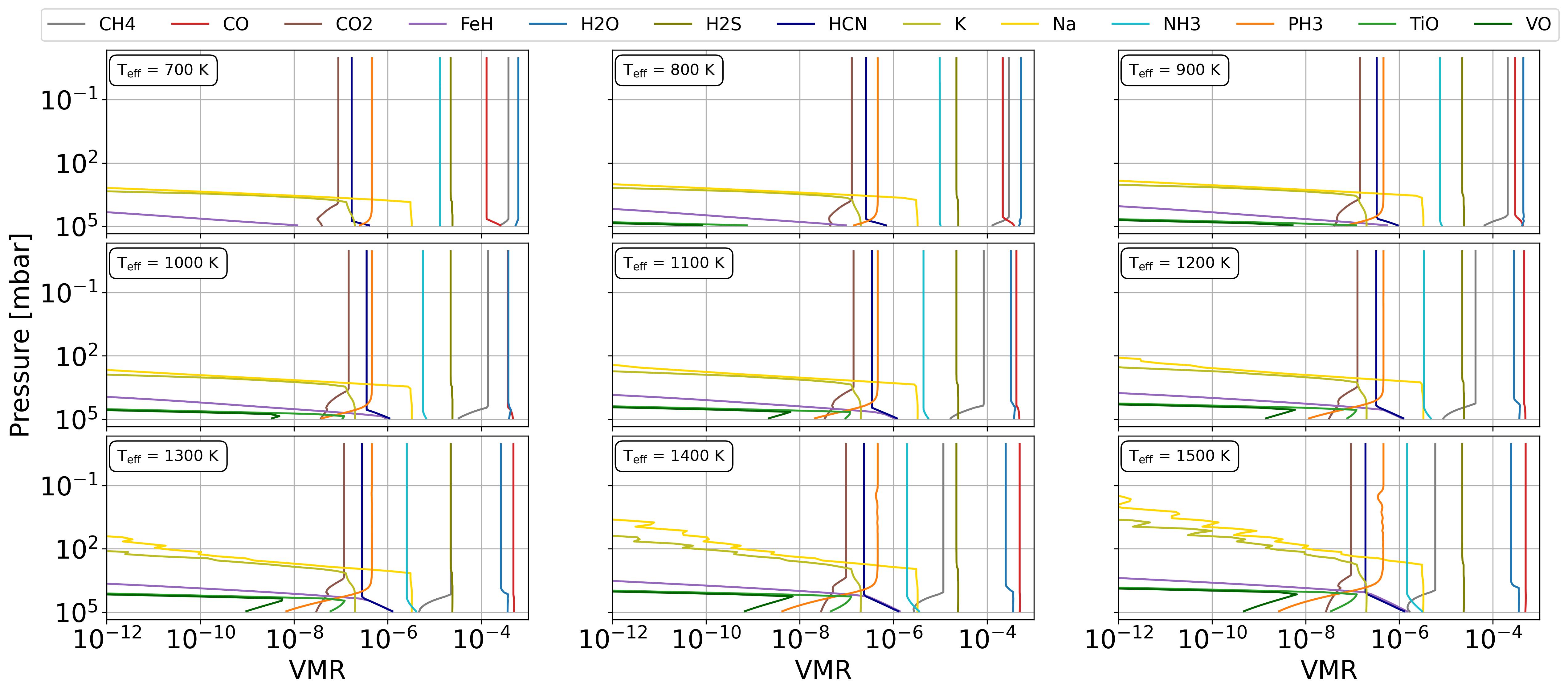}
    \caption{Vertical abundance profiles of the considered species obtained using \texttt{Exo-REM} for each effective temperature used in this work. The effective temperature rises from the upper left corner to the bottom right. We observe a transition in the dominant chemical species from a H$_{\rm 2}$ O and CH$_{\rm 4}$ dominated atmosphere to a CO-dominated atmosphere, as the atmosphere warms. The transition occurs for an effective temperature of 1000 K. TiO and VO are included in the calculations although they are negligible for low effective temperature, and only have a minor effect on the chemical structure at higher effective temperatures.}
    \label{fig:sup1}
\end{figure}
\clearpage
\bibliography{references}

\begin{thebibliography}{10}
\expandafter\ifx\csname url\endcsname\relax
  \def\url#1{\burl{#1}}\fi
\expandafter\ifx\csname urlprefix\endcsname\relax\def\urlprefix{URL }\fi
\providecommand{\bibinfo}[2]{#2}
\providecommand{\eprint}[2][]{\url{#2}}
\providecommand{\doi}[1]{\url{https://doi.org/#1}}
\bibcommenthead

\bibitem{kirkpatrick_new_2005}
\bibinfo{author}{Kirkpatrick, J.~D.}
\newblock \bibinfo{title}{New {Spectral} {Types} {L} and {T}}.
\newblock \emph{\bibinfo{journal}{Annual Review of Astronomy and Astrophysics}} \textbf{\bibinfo{volume}{43}}, \bibinfo{pages}{195--245} (\bibinfo{year}{2005}).
\newblock \urlprefix\url{https://ui.adsabs.harvard.edu/abs/2005ARA&A..43..195K/abstract}.

\bibitem{cushing_discovery_2011}
\bibinfo{author}{Cushing, M.~C.} \emph{et~al.}
\newblock \bibinfo{title}{The {Discovery} of {Y} {Dwarfs} using {Data} from the {Wide}-field {Infrared} {Survey} {Explorer} ({WISE})}.
\newblock \emph{\bibinfo{journal}{The Astrophysical Journal}} \textbf{\bibinfo{volume}{743}}, \bibinfo{pages}{50} (\bibinfo{year}{2011}).
\newblock \urlprefix\url{https://ui.adsabs.harvard.edu/abs/2011ApJ...743...50C}.
\newblock \bibinfo{note}{Publisher: IOP ADS Bibcode: 2011ApJ...743...50C}.

\bibitem{vos_astro2020_2019}
\bibinfo{author}{Vos, J.~M.} \emph{et~al.}
\newblock \bibinfo{title}{Astro2020 {White} {Paper}: {The} {L}/{T} {Transition}}.
\newblock \bibinfo{type}{Tech. Rep.} (\bibinfo{year}{2019}).
\newblock \urlprefix\url{https://ui.adsabs.harvard.edu/abs/2019arXiv190306691V}.
\newblock \bibinfo{note}{Publication Title: arXiv e-prints ADS Bibcode: 2019arXiv190306691V Type: article}.

\bibitem{tsuji_warm_1999}
\bibinfo{author}{Tsuji, T.}, \bibinfo{author}{Ohnaka, K.} \& \bibinfo{author}{Aoki, W.}
\newblock \bibinfo{title}{Warm {Dust} in the {Cool} {Brown} {Dwarf} {Gliese} {229B} and {Spectroscopic} {Diagnosis} of {Dusty} {Photospheres}}.
\newblock \emph{\bibinfo{journal}{The Astrophysical Journal}} \textbf{\bibinfo{volume}{520}}, \bibinfo{pages}{L119--L122} (\bibinfo{year}{1999}).
\newblock \urlprefix\url{https://ui.adsabs.harvard.edu/abs/1999ApJ...520L.119T}.
\newblock \bibinfo{note}{Publisher: IOP ADS Bibcode: 1999ApJ...520L.119T}.

\bibitem{allard_limiting_2001}
\bibinfo{author}{Allard, F.}, \bibinfo{author}{Hauschildt, P.~H.}, \bibinfo{author}{Alexander, D.~R.}, \bibinfo{author}{Tamanai, A.} \& \bibinfo{author}{Schweitzer, A.}
\newblock \bibinfo{title}{The {Limiting} {Effects} of {Dust} in {Brown} {Dwarf} {Model} {Atmospheres}}.
\newblock \emph{\bibinfo{journal}{The Astrophysical Journal, Volume 556, Issue 1, pp. 357-372.}} \textbf{\bibinfo{volume}{556}}, \bibinfo{pages}{357} (\bibinfo{year}{2001}).
\newblock \urlprefix\url{https://ui.adsabs.harvard.edu/abs/2001ApJ...556..357A/abstract}.

\bibitem{ackerman_precipitating_2001}
\bibinfo{author}{Ackerman, A.~S.} \& \bibinfo{author}{Marley, M.~S.}
\newblock \bibinfo{title}{Precipitating {Condensation} {Clouds} in {Substellar} {Atmospheres}}.
\newblock \emph{\bibinfo{journal}{The Astrophysical Journal}} \textbf{\bibinfo{volume}{556}}, \bibinfo{pages}{872--884} (\bibinfo{year}{2001}).
\newblock \urlprefix\url{https://ui.adsabs.harvard.edu/abs/2001ApJ...556..872A}.
\newblock \bibinfo{note}{ADS Bibcode: 2001ApJ...556..872A}.

\bibitem{helling_atmospheres_2014}
\bibinfo{author}{Helling, C.} \& \bibinfo{author}{Casewell, S.}
\newblock \bibinfo{title}{Atmospheres of {Brown} {Dwarfs}}.
\newblock \emph{\bibinfo{journal}{The Astronomy and Astrophysics Review}} \textbf{\bibinfo{volume}{22}}, \bibinfo{pages}{80} (\bibinfo{year}{2014}).
\newblock \urlprefix\url{http://arxiv.org/abs/1410.6029}.
\newblock \bibinfo{note}{ArXiv:1410.6029 [astro-ph, physics:physics]}.

\bibitem{charnay_self-consistent_2018}
\bibinfo{author}{Charnay, B.} \emph{et~al.}
\newblock \bibinfo{title}{A {Self}-consistent {Cloud} {Model} for {Brown} {Dwarfs} and {Young} {Giant} {Exoplanets}: {Comparison} with {Photometric} and {Spectroscopic} {Observations}}.
\newblock \emph{\bibinfo{journal}{The Astrophysical Journal}} \textbf{\bibinfo{volume}{854}}, \bibinfo{pages}{172} (\bibinfo{year}{2018}).
\newblock \urlprefix\url{https://iopscience.iop.org/article/10.3847/1538-4357/aaac7d}.

\bibitem{saumon_evolution_2008}
\bibinfo{author}{Saumon, D.} \& \bibinfo{author}{Marley, M.~S.}
\newblock \bibinfo{title}{The {Evolution} of {L} and {T} {Dwarfs} in {Color}-{Magnitude} {Diagrams}}.
\newblock \emph{\bibinfo{journal}{The Astrophysical Journal}} \textbf{\bibinfo{volume}{689}}, \bibinfo{pages}{1327--1344} (\bibinfo{year}{2008}).
\newblock \urlprefix\url{https://ui.adsabs.harvard.edu/abs/2008ApJ...689.1327S}.
\newblock \bibinfo{note}{Publisher: IOP ADS Bibcode: 2008ApJ...689.1327S}.

\bibitem{allard_models_2012}
\bibinfo{author}{Allard, F.}, \bibinfo{author}{Homeier, D.} \& \bibinfo{author}{Freytag, B.}
\newblock \bibinfo{title}{Models of very-low-mass stars, brown dwarfs and exoplanets}.
\newblock \emph{\bibinfo{journal}{Philosophical Transactions of the Royal Society of London Series A}} \textbf{\bibinfo{volume}{370}}, \bibinfo{pages}{2765--2777} (\bibinfo{year}{2012}).
\newblock \urlprefix\url{https://ui.adsabs.harvard.edu/abs/2012RSPTA.370.2765A}.
\newblock \bibinfo{note}{ADS Bibcode: 2012RSPTA.370.2765A}.

\bibitem{marley_patchy_2010}
\bibinfo{author}{Marley, M.~S.}, \bibinfo{author}{Saumon, D.} \& \bibinfo{author}{Goldblatt, C.}
\newblock \bibinfo{title}{A {Patchy} {Cloud} {Model} for the {L} to {T} {Dwarf} {Transition}}.
\newblock \emph{\bibinfo{journal}{The Astrophysical Journal}} \textbf{\bibinfo{volume}{723}}, \bibinfo{pages}{L117--L121} (\bibinfo{year}{2010}).
\newblock \urlprefix\url{https://ui.adsabs.harvard.edu/abs/2010ApJ...723L.117M}.
\newblock \bibinfo{note}{Publisher: IOP ADS Bibcode: 2010ApJ...723L.117M}.

\bibitem{tremblin_thermo-compositional_2019}
\bibinfo{author}{Tremblin, P.} \emph{et~al.}
\newblock \bibinfo{title}{Thermo-compositional {Diabatic} {Convection} in the {Atmospheres} of {Brown} {Dwarfs} and in {Earth}’s {Atmosphere} and {Oceans}}.
\newblock \emph{\bibinfo{journal}{The Astrophysical Journal}} \textbf{\bibinfo{volume}{876}}, \bibinfo{pages}{144} (\bibinfo{year}{2019}).
\newblock \urlprefix\url{https://ui.adsabs.harvard.edu/abs/2019ApJ...876..144T}.
\newblock \bibinfo{note}{ADS Bibcode: 2019ApJ...876..144T}.

\bibitem{tremblin_rotational_2020}
\bibinfo{author}{Tremblin, P.} \emph{et~al.}
\newblock \bibinfo{title}{Rotational spectral modulation of cloudless atmospheres for {L}/{T} brown dwarfs and extrasolar giant planets}.
\newblock \emph{\bibinfo{journal}{Astronomy and Astrophysics}} \textbf{\bibinfo{volume}{643}}, \bibinfo{pages}{A23} (\bibinfo{year}{2020}).
\newblock \urlprefix\url{https://ui.adsabs.harvard.edu/abs/2020A&A...643A..23T/abstract}.

\bibitem{metchev_weather_2015}
\bibinfo{author}{Metchev, S.~A.} \emph{et~al.}
\newblock \bibinfo{title}{{WEATHER} {ON} {OTHER} {WORLDS}. {II}. {SURVEY} {RESULTS}: {SPOTS} {ARE} {UBIQUITOUS} {ON} {L} {AND} {T} {DWARFS}}.
\newblock \emph{\bibinfo{journal}{The Astrophysical Journal}} \textbf{\bibinfo{volume}{799}}, \bibinfo{pages}{154} (\bibinfo{year}{2015}).
\newblock \urlprefix\url{https://iopscience.iop.org/article/10.1088/0004-637X/799/2/154}.

\bibitem{crossfield_global_2014}
\bibinfo{author}{Crossfield, I. J.~M.} \emph{et~al.}
\newblock \bibinfo{title}{A global cloud map of the nearest known brown dwarf}.
\newblock \emph{\bibinfo{journal}{Nature}} \textbf{\bibinfo{volume}{505}}, \bibinfo{pages}{654--656} (\bibinfo{year}{2014}).
\newblock \urlprefix\url{https://ui.adsabs.harvard.edu/abs/2014Natur.505..654C}.
\newblock \bibinfo{note}{ADS Bibcode: 2014Natur.505..654C}.

\bibitem{artigau_photometric_2009}
\bibinfo{author}{Artigau, E.}, \bibinfo{author}{Bouchard, S.}, \bibinfo{author}{Doyon, R.} \& \bibinfo{author}{Lafreniere, D.}
\newblock \bibinfo{title}{Photometric {Variability} of the {T2}.5 {Brown} {Dwarf} {SIMP} {J013656}.5+093347: {Evidence} for {Evolving} {Weather} {Patterns}}.
\newblock \emph{\bibinfo{journal}{The Astrophysical Journal}} \textbf{\bibinfo{volume}{701}}, \bibinfo{pages}{1534--1539} (\bibinfo{year}{2009}).
\newblock \urlprefix\url{https://ui.adsabs.harvard.edu/abs/2009ApJ...701.1534A}.
\newblock \bibinfo{note}{Publisher: IOP ADS Bibcode: 2009ApJ...701.1534A}.

\bibitem{radigan_strong_2014}
\bibinfo{author}{Radigan, J.}, \bibinfo{author}{Lafrenière, D.}, \bibinfo{author}{Jayawardhana, R.} \& \bibinfo{author}{Artigau, E.}
\newblock \bibinfo{title}{Strong {Brightness} {Variations} {Signal} {Cloudy}-to-clear {Transition} of {Brown} {Dwarfs}}.
\newblock \emph{\bibinfo{journal}{The Astrophysical Journal}} \textbf{\bibinfo{volume}{793}}, \bibinfo{pages}{75} (\bibinfo{year}{2014}).
\newblock \urlprefix\url{https://ui.adsabs.harvard.edu/abs/2014ApJ...793...75R}.
\newblock \bibinfo{note}{Publisher: IOP ADS Bibcode: 2014ApJ...793...75R}.

\bibitem{apai_hst_2013}
\bibinfo{author}{Apai, D.} \emph{et~al.}
\newblock \bibinfo{title}{\textit{{HST}} {SPECTRAL} {MAPPING} {OF} {L}/{T} {TRANSITION} {BROWN} {DWARFS} {REVEALS} {CLOUD} {THICKNESS} {VARIATIONS}}.
\newblock \emph{\bibinfo{journal}{The Astrophysical Journal}} \textbf{\bibinfo{volume}{768}}, \bibinfo{pages}{121} (\bibinfo{year}{2013}).
\newblock \urlprefix\url{https://iopscience.iop.org/article/10.1088/0004-637X/768/2/121}.

\bibitem{buenzli_vertical_2012}
\bibinfo{author}{Buenzli, E.} \emph{et~al.}
\newblock \bibinfo{title}{{VERTICAL} {ATMOSPHERIC} {STRUCTURE} {IN} {A} {VARIABLE} {BROWN} {DWARF}: {PRESSURE}-{DEPENDENT} {PHASE} {SHIFTS} {IN} {SIMULTANEOUS} \textit{{HUBBLE} {SPACE} {TELESCOPE}} - \textit{{SPITZER}} {LIGHT} {CURVES}}.
\newblock \emph{\bibinfo{journal}{The Astrophysical Journal}} \textbf{\bibinfo{volume}{760}}, \bibinfo{pages}{L31} (\bibinfo{year}{2012}).
\newblock \urlprefix\url{https://iopscience.iop.org/article/10.1088/2041-8205/760/2/L31}.

\bibitem{buenzli_cloud_2015}
\bibinfo{author}{Buenzli, E.} \emph{et~al.}
\newblock \bibinfo{title}{{CLOUD} {STRUCTURE} {OF} {THE} {NEAREST} {BROWN} {DWARFS}: {SPECTROSCOPIC} {VARIABILITY} {OF} {LUHMAN} {16AB} {FROM} {THE} \textit{{HUBBLE} {SPACE} {TELESCOPE}}}.
\newblock \emph{\bibinfo{journal}{The Astrophysical Journal}} \textbf{\bibinfo{volume}{798}}, \bibinfo{pages}{127} (\bibinfo{year}{2015}).
\newblock \urlprefix\url{https://iopscience.iop.org/article/10.1088/0004-637X/798/2/127}.

\bibitem{apai_zones_2017}
\bibinfo{author}{Apai, D.} \emph{et~al.}
\newblock \bibinfo{title}{Zones, spots, and planetary-scale waves beating in brown dwarf atmospheres}.
\newblock \emph{\bibinfo{journal}{Science}} \textbf{\bibinfo{volume}{357}}, \bibinfo{pages}{683--687} (\bibinfo{year}{2017}).
\newblock \urlprefix\url{https://www.science.org/doi/10.1126/science.aam9848}.
\newblock \bibinfo{note}{Publisher: American Association for the Advancement of Science}.

\bibitem{robinson_temperature_2014}
\bibinfo{author}{Robinson, T.~D.} \& \bibinfo{author}{Marley, M.~S.}
\newblock \bibinfo{title}{Temperature {Fluctuations} as a {Source} of {Brown} {Dwarf} {Variability}}.
\newblock \emph{\bibinfo{journal}{The Astrophysical Journal}} \textbf{\bibinfo{volume}{785}}, \bibinfo{pages}{158} (\bibinfo{year}{2014}).
\newblock \urlprefix\url{https://ui.adsabs.harvard.edu/abs/2014ApJ...785..158R}.
\newblock \bibinfo{note}{Publisher: IOP ADS Bibcode: 2014ApJ...785..158R}.

\bibitem{morley_water_2014}
\bibinfo{author}{Morley, C.~V.} \emph{et~al.}
\newblock \bibinfo{title}{Water {Clouds} in {Y} {Dwarfs} and {Exoplanets}}.
\newblock \emph{\bibinfo{journal}{The Astrophysical Journal}} \textbf{\bibinfo{volume}{787}}, \bibinfo{pages}{78} (\bibinfo{year}{2014}).
\newblock \urlprefix\url{https://ui.adsabs.harvard.edu/abs/2014ApJ...787...78M}.
\newblock \bibinfo{note}{ADS Bibcode: 2014ApJ...787...78M}.

\bibitem{tan_atmospheric_2019}
\bibinfo{author}{Tan, X.} \& \bibinfo{author}{Showman, A.~P.}
\newblock \bibinfo{title}{Atmospheric {Variability} {Driven} by {Radiative} {Cloud} {Feedback} in {Brown} {Dwarfs} and {Directly} {Imaged} {Extrasolar} {Giant} {Planets}}.
\newblock \emph{\bibinfo{journal}{The Astrophysical Journal}} \textbf{\bibinfo{volume}{874}}, \bibinfo{pages}{111} (\bibinfo{year}{2019}).
\newblock \urlprefix\url{https://iopscience.iop.org/article/10.3847/1538-4357/ab0c07}.

\bibitem{luna_empirically_2021}
\bibinfo{author}{Luna, J.~L.} \& \bibinfo{author}{Morley, C.~V.}
\newblock \bibinfo{title}{Empirically {Determining} {Substellar} {Cloud} {Compositions} in the {Era} of the {James} {Webb} {Space} {Telescope}}.
\newblock \emph{\bibinfo{journal}{The Astrophysical Journal}} \textbf{\bibinfo{volume}{920}}, \bibinfo{pages}{146} (\bibinfo{year}{2021}).
\newblock \urlprefix\url{https://ui.adsabs.harvard.edu/abs/2021ApJ...920..146L}.
\newblock \bibinfo{note}{Publisher: IOP ADS Bibcode: 2021ApJ...920..146L}.

\bibitem{showman_atmospheric_2013}
\bibinfo{author}{Showman, A.~P.} \& \bibinfo{author}{Kaspi, Y.}
\newblock \bibinfo{title}{Atmospheric {Dynamics} of {Brown} {Dwarfs} and {Directly} {Imaged} {Giant} {Planets}}.
\newblock \emph{\bibinfo{journal}{The Astrophysical Journal}} \textbf{\bibinfo{volume}{776}}, \bibinfo{pages}{85} (\bibinfo{year}{2013}).
\newblock \urlprefix\url{https://ui.adsabs.harvard.edu/abs/2013ApJ...776...85S}.
\newblock \bibinfo{note}{Publisher: IOP ADS Bibcode: 2013ApJ...776...85S}.

\bibitem{zhang_atmospheric_2014}
\bibinfo{author}{Zhang, X.} \& \bibinfo{author}{Showman, A.~P.}
\newblock \bibinfo{title}{Atmospheric {Circulation} of {Brown} {Dwarfs}: {Jets}, {Vortices}, and {Time} {Variability}}.
\newblock \emph{\bibinfo{journal}{The Astrophysical Journal}} \textbf{\bibinfo{volume}{788}}, \bibinfo{pages}{L6} (\bibinfo{year}{2014}).
\newblock \urlprefix\url{https://ui.adsabs.harvard.edu/abs/2014ApJ...788L...6Z}.
\newblock \bibinfo{note}{Publisher: IOP ADS Bibcode: 2014ApJ...788L...6Z}.

\bibitem{showman_atmospheric_2019}
\bibinfo{author}{Showman, A.~P.}, \bibinfo{author}{Tan, X.} \& \bibinfo{author}{Zhang, X.}
\newblock \bibinfo{title}{Atmospheric {Circulation} of {Brown} {Dwarfs} and {Jupiter}- and {Saturn}-like {Planets}: {Zonal} {Jets}, {Long}-term {Variability}, and {QBO}-type {Oscillations}}.
\newblock \emph{\bibinfo{journal}{The Astrophysical Journal}} \textbf{\bibinfo{volume}{883}}, \bibinfo{pages}{4} (\bibinfo{year}{2019}).
\newblock \urlprefix\url{https://iopscience.iop.org/article/10.3847/1538-4357/ab384a}.

\bibitem{tan_jet_2022}
\bibinfo{author}{Tan, X.}
\newblock \bibinfo{title}{Jet streams and tracer mixing in the atmospheres of brown dwarfs and isolated young giant planets}.
\newblock \emph{\bibinfo{journal}{Monthly Notices of the Royal Astronomical Society}} \textbf{\bibinfo{volume}{511}}, \bibinfo{pages}{4861--4881} (\bibinfo{year}{2022}).
\newblock \urlprefix\url{https://academic.oup.com/mnras/article/511/4/4861/6524914}.

\bibitem{lee_dynamically_2023}
\bibinfo{author}{Lee, E. K.~H.}, \bibinfo{author}{Tan, X.} \& \bibinfo{author}{Tsai, S.-M.}
\newblock \bibinfo{title}{Dynamically coupled kinetic chemistry in brown dwarf atmospheres {I}. {Performing} global scale kinetic modelling} (\bibinfo{year}{2023}).
\newblock \urlprefix\url{http://arxiv.org/abs/2306.03520}.
\newblock \bibinfo{note}{ArXiv:2306.03520 [astro-ph]}.

\bibitem{tan_atmospheric_2021-1}
\bibinfo{author}{Tan, X.} \& \bibinfo{author}{Showman, A.~P.}
\newblock \bibinfo{title}{Atmospheric circulation of brown dwarfs and directly imaged exoplanets driven by cloud radiative feedback: effects of rotation}.
\newblock \emph{\bibinfo{journal}{Monthly Notices of the Royal Astronomical Society}} \textbf{\bibinfo{volume}{502}}, \bibinfo{pages}{678--699} (\bibinfo{year}{2021}).
\newblock \urlprefix\url{https://academic.oup.com/mnras/article/502/1/678/6089155}.

\bibitem{freytag_role_2010}
\bibinfo{author}{Freytag, B.}, \bibinfo{author}{Allard, F.}, \bibinfo{author}{Ludwig, H.~G.}, \bibinfo{author}{Homeier, D.} \& \bibinfo{author}{Steffen, M.}
\newblock \bibinfo{title}{The role of convection, overshoot, and gravity waves for the transport of dust in {M} dwarf and brown dwarf atmospheres}.
\newblock \emph{\bibinfo{journal}{Astronomy and Astrophysics}} \textbf{\bibinfo{volume}{513}}, \bibinfo{pages}{A19} (\bibinfo{year}{2010}).
\newblock \urlprefix\url{https://ui.adsabs.harvard.edu/abs/2010A&A...513A..19F}.
\newblock \bibinfo{note}{ADS Bibcode: 2010A\&A...513A..19F}.

\bibitem{lefevre_cloud-convection_2022}
\bibinfo{author}{Lefèvre, M.}, \bibinfo{author}{Tan, X.}, \bibinfo{author}{Lee, E. K.~H.} \& \bibinfo{author}{Pierrehumbert, R.~T.}
\newblock \bibinfo{title}{Cloud-convection {Feedback} in {Brown} {Dwarf} {Atmospheres}}.
\newblock \emph{\bibinfo{journal}{The Astrophysical Journal}} \textbf{\bibinfo{volume}{929}}, \bibinfo{pages}{153} (\bibinfo{year}{2022}).
\newblock \urlprefix\url{https://iopscience.iop.org/article/10.3847/1538-4357/ac5e2d}.

\bibitem{wordsworth_gliese_2011}
\bibinfo{author}{Wordsworth, R.~D.} \emph{et~al.}
\newblock \bibinfo{title}{Gliese 581d is the {First} {Discovered} {Terrestrial}-mass {Exoplanet} in the {Habitable} {Zone}}.
\newblock \emph{\bibinfo{journal}{The Astrophysical Journal}} \textbf{\bibinfo{volume}{733}}, \bibinfo{pages}{L48} (\bibinfo{year}{2011}).
\newblock \urlprefix\url{https://ui.adsabs.harvard.edu/abs/2011ApJ...733L..48W}.
\newblock \bibinfo{note}{Publisher: IOP ADS Bibcode: 2011ApJ...733L..48W}.

\bibitem{teinturier_radiative_2024}
\bibinfo{author}{Teinturier, L.} \emph{et~al.}
\newblock \bibinfo{title}{The radiative and dynamical impact of clouds in the atmosphere of the hot {Jupiter} {WASP}-43 b}.
\newblock \emph{\bibinfo{journal}{Astronomy \& Astrophysics}} \textbf{\bibinfo{volume}{683}}, \bibinfo{pages}{A231} (\bibinfo{year}{2024}).
\newblock \urlprefix\url{https://www.aanda.org/10.1051/0004-6361/202347069}.

\bibitem{visscher_atmospheric_2010}
\bibinfo{author}{Visscher, C.}, \bibinfo{author}{Lodders, K.} \& \bibinfo{author}{Fegley, B.}
\newblock \bibinfo{title}{{ATMOSPHERIC} {CHEMISTRY} {IN} {GIANT} {PLANETS}, {BROWN} {DWARFS}, {AND} {LOW}-{MASS} {DWARF} {STARS}. {III}. {IRON}, {MAGNESIUM}, {AND} {SILICON}}.
\newblock \emph{\bibinfo{journal}{The Astrophysical Journal}} \textbf{\bibinfo{volume}{716}}, \bibinfo{pages}{1060--1075} (\bibinfo{year}{2010}).
\newblock \urlprefix\url{https://iopscience.iop.org/article/10.1088/0004-637X/716/2/1060}.

\bibitem{showman_atmospheric_2020}
\bibinfo{author}{Showman, A.~P.}, \bibinfo{author}{Tan, X.} \& \bibinfo{author}{Parmentier, V.}
\newblock \bibinfo{title}{Atmospheric {Dynamics} of {Hot} {Giant} {Planets} and {Brown} {Dwarfs}}.
\newblock \emph{\bibinfo{journal}{Space Science Reviews}} \textbf{\bibinfo{volume}{216}}, \bibinfo{pages}{139} (\bibinfo{year}{2020}).
\newblock \urlprefix\url{http://arxiv.org/abs/2007.15363}.
\newblock \bibinfo{note}{ArXiv:2007.15363 [astro-ph]}.

\bibitem{tan_atmospheric_2021}
\bibinfo{author}{Tan, X.} \& \bibinfo{author}{Showman, A.~P.}
\newblock \bibinfo{title}{Atmospheric circulation of brown dwarfs and directly imaged exoplanets driven by cloud radiative feedback: global and equatorial dynamics}.
\newblock \emph{\bibinfo{journal}{Monthly Notices of the Royal Astronomical Society}} \textbf{\bibinfo{volume}{502}}, \bibinfo{pages}{2198--2219} (\bibinfo{year}{2021}).
\newblock \urlprefix\url{https://academic.oup.com/mnras/article/502/2/2198/6101225}.

\bibitem{lee_dynamically_2024}
\bibinfo{author}{Lee, E. K.~H.}, \bibinfo{author}{Tan, X.} \& \bibinfo{author}{Tsai, S.-M.}
\newblock \bibinfo{title}{Dynamically coupled kinetic chemistry in brown dwarf atmospheres – {II}. {Cloud} and chemistry connections in directly imaged sub-{Jupiter} exoplanets}.
\newblock \emph{\bibinfo{journal}{Monthly Notices of the Royal Astronomical Society}} \textbf{\bibinfo{volume}{529}}, \bibinfo{pages}{2686--2701} (\bibinfo{year}{2024}).
\newblock \urlprefix\url{https://academic.oup.com/mnras/article/529/3/2686/7611715}.

\bibitem{allers_measurement_2020}
\bibinfo{author}{Allers, K.~N.}, \bibinfo{author}{Vos, J.~M.}, \bibinfo{author}{Biller, B.~A.} \& \bibinfo{author}{Williams, P. K.~G.}
\newblock \bibinfo{title}{A measurement of the wind speed on a brown dwarf}.
\newblock \emph{\bibinfo{journal}{Science}} \textbf{\bibinfo{volume}{368}}, \bibinfo{pages}{169--172} (\bibinfo{year}{2020}).
\newblock \urlprefix\url{https://www.science.org/doi/10.1126/science.aaz2856}.
\newblock \bibinfo{note}{Publisher: American Association for the Advancement of Science}.

\bibitem{vos_viewing_2017}
\bibinfo{author}{Vos, J.~M.}, \bibinfo{author}{Allers, K.~N.} \& \bibinfo{author}{Biller, B.~A.}
\newblock \bibinfo{title}{The {Viewing} {Geometry} of {Brown} {Dwarfs} {Influences} {Their} {Observed} {Colors} and {Variability} {Amplitudes}}.
\newblock \emph{\bibinfo{journal}{The Astrophysical Journal}} \textbf{\bibinfo{volume}{842}}, \bibinfo{pages}{78} (\bibinfo{year}{2017}).
\newblock \urlprefix\url{https://iopscience.iop.org/article/10.3847/1538-4357/aa73cf}.

\bibitem{suarez_ultracool_2023}
\bibinfo{author}{Suárez, G.}, \bibinfo{author}{Vos, J.~M.}, \bibinfo{author}{Metchev, S.}, \bibinfo{author}{Faherty, J.~K.} \& \bibinfo{author}{Cruz, K.}
\newblock \bibinfo{title}{Ultracool {Dwarfs} {Observed} with the {Spitzer} {Infrared} {Spectrograph}: {Equatorial} {Latitudes} in {L} {Dwarf} {Atmospheres} {Are} {Cloudier}}.
\newblock \emph{\bibinfo{journal}{The Astrophysical Journal Letters}} \textbf{\bibinfo{volume}{954}}, \bibinfo{pages}{L6} (\bibinfo{year}{2023}).
\newblock \urlprefix\url{https://dx.doi.org/10.3847/2041-8213/acec4b}.
\newblock \bibinfo{note}{Publisher: The American Astronomical Society}.

\bibitem{vos_variability_2018}
\bibinfo{author}{Vos, J.~M.} \emph{et~al.}
\newblock \bibinfo{title}{Variability of the lowest mass objects in the {AB} {Doradus} moving group}.
\newblock \emph{\bibinfo{journal}{Monthly Notices of the Royal Astronomical Society}} \textbf{\bibinfo{volume}{474}}, \bibinfo{pages}{1041--1053} (\bibinfo{year}{2018}).
\newblock \urlprefix\url{https://doi.org/10.1093/mnras/stx2752}.

\bibitem{fuda_latitude-dependent_2024}
\bibinfo{author}{Fuda, N.} \emph{et~al.}
\newblock \bibinfo{title}{Latitude-dependent {Atmospheric} {Waves} and {Long}-period {Modulations} in {Luhman} 16 {B} from the {Longest} {Lightcurve} of an {Extrasolar} {World}} (\bibinfo{year}{2024}).
\newblock \urlprefix\url{http://arxiv.org/abs/2403.02260}.
\newblock \bibinfo{note}{ArXiv:2403.02260 null}.

\bibitem{sardeshmukh_generation_1988}
\bibinfo{author}{Sardeshmukh, P.~D.} \& \bibinfo{author}{Hoskins, B.~J.}
\newblock \bibinfo{title}{The {Generation} of {Global} {Rotational} {Flow} by {Steady} {Idealized} {Tropical} {Divergence}.}
\newblock \emph{\bibinfo{journal}{Journal of the Atmospheric Sciences}} \textbf{\bibinfo{volume}{45}}, \bibinfo{pages}{1228--1251} (\bibinfo{year}{1988}).
\newblock \urlprefix\url{https://ui.adsabs.harvard.edu/abs/1988JAtS...45.1228S}.
\newblock \bibinfo{note}{ADS Bibcode: 1988JAtS...45.1228S}.

\bibitem{schneider_formation_2009}
\bibinfo{author}{Schneider, T.} \& \bibinfo{author}{Liu, J.}
\newblock \bibinfo{title}{Formation of {Jets} and {Equatorial} {Superrotation} on {Jupiter}}.
\newblock \emph{\bibinfo{journal}{Journal of the Atmospheric Sciences}} \textbf{\bibinfo{volume}{66}}, \bibinfo{pages}{579--601} (\bibinfo{year}{2009}).
\newblock \urlprefix\url{http://arxiv.org/abs/0809.4302}.
\newblock \bibinfo{note}{ArXiv:0809.4302 [astro-ph, physics:physics]}.

\bibitem{liu_mechanisms_2010}
\bibinfo{author}{Liu, J.} \& \bibinfo{author}{Schneider, T.}
\newblock \bibinfo{title}{Mechanisms of jet formation on the giant planets}.
\newblock \emph{\bibinfo{journal}{Journal of the Atmospheric Sciences}} \textbf{\bibinfo{volume}{67}}, \bibinfo{pages}{3652--3672} (\bibinfo{year}{2010}).
\newblock \urlprefix\url{http://arxiv.org/abs/0910.3682}.
\newblock \bibinfo{note}{ArXiv:0910.3682 [astro-ph]}.

\bibitem{spiga_global_2020}
\bibinfo{author}{Spiga, A.} \emph{et~al.}
\newblock \bibinfo{title}{Global climate modeling of {Saturn}'s atmosphere. {Part} {II}: {Multi}-annual high-resolution dynamical simulations}.
\newblock \emph{\bibinfo{journal}{Icarus}} \textbf{\bibinfo{volume}{335}}, \bibinfo{pages}{113377} (\bibinfo{year}{2020}).
\newblock \urlprefix\url{https://ui.adsabs.harvard.edu/abs/2020Icar..33513377S}.
\newblock \bibinfo{note}{ADS Bibcode: 2020Icar..33513377S}.

\bibitem{wheeler_convectively_1999}
\bibinfo{author}{Wheeler, M.} \& \bibinfo{author}{Kiladis, G.~N.}
\newblock \bibinfo{title}{Convectively {Coupled} {Equatorial} {Waves}: {Analysis} of {Clouds} and {Temperature} in the {Wavenumber}-{Frequency} {Domain}.}
\newblock \emph{\bibinfo{journal}{Journal of the Atmospheric Sciences}} \textbf{\bibinfo{volume}{56}}, \bibinfo{pages}{374--399} (\bibinfo{year}{1999}).
\newblock \urlprefix\url{https://ui.adsabs.harvard.edu/abs/1999JAtS...56..374W}.
\newblock \bibinfo{note}{ADS Bibcode: 1999JAtS...56..374W}.

\bibitem{yang_extrasolar_2016}
\bibinfo{author}{Yang, H.} \emph{et~al.}
\newblock \bibinfo{title}{{EXTRASOLAR} {STORMS}: {PRESSURE}-{DEPENDENT} {CHANGES} {IN} {LIGHT}-{CURVE} {PHASE} {IN} {BROWN} {DWARFS} {FROM} {SIMULTANEOUS} {HST} {AND} {SPITZER} {OBSERVATIONS}}.
\newblock \emph{\bibinfo{journal}{The Astrophysical Journal}} \textbf{\bibinfo{volume}{826}}, \bibinfo{pages}{8} (\bibinfo{year}{2016}).
\newblock \urlprefix\url{https://iopscience.iop.org/article/10.3847/0004-637X/826/1/8}.

\bibitem{artigau_variability_2018}
\bibinfo{author}{Artigau, E.}
\newblock \emph{\bibinfo{title}{Variability of {Brown} {Dwarfs}}}  (\bibinfo{year}{2018}).
\newblock \urlprefix\url{https://ui.adsabs.harvard.edu/abs/2018haex.bookE..94A}.
\newblock \bibinfo{note}{Pages: 94 Publication Title: Handbook of Exoplanets ADS Bibcode: 2018haex.bookE..94A}.

\bibitem{biller_clouds_2023}
\bibinfo{author}{Biller, B.} \emph{et~al.}
\newblock \bibinfo{title}{Clouds or {Chemistry}?: {Pinpointing} the drivers of variability across the {L}/{T} transition via the benchmark {L}/{T} binary {WISE} {1049AB}}.
\newblock \emph{\bibinfo{journal}{JWST Proposal. Cycle 2}} \bibinfo{pages}{2965} (\bibinfo{year}{2023}).
\newblock \urlprefix\url{https://ui.adsabs.harvard.edu/abs/2023jwst.prop.2965B}.
\newblock \bibinfo{note}{ADS Bibcode: 2023jwst.prop.2965B}.

\bibitem{plummer_mapping_2023}
\bibinfo{author}{Plummer, M.~K.} \& \bibinfo{author}{Wang, J.}
\newblock \bibinfo{title}{Mapping the {Skies} of {Ultracool} {Worlds}: {Detecting} {Storms} and {Spots} with {Extremely} {Large} {Telescopes}}.
\newblock \emph{\bibinfo{journal}{The Astrophysical Journal}} \textbf{\bibinfo{volume}{951}}, \bibinfo{pages}{101} (\bibinfo{year}{2023}).
\newblock \urlprefix\url{https://dx.doi.org/10.3847/1538-4357/accd5d}.
\newblock \bibinfo{note}{Publisher: The American Astronomical Society}.

\bibitem{schneider_possible_2019}
\bibinfo{author}{Schneider, T.}, \bibinfo{author}{Kaul, C.~M.} \& \bibinfo{author}{Pressel, K.~G.}
\newblock \bibinfo{title}{Possible climate transitions from breakup of stratocumulus decks under greenhouse warming}.
\newblock \emph{\bibinfo{journal}{Nature Geoscience}} \textbf{\bibinfo{volume}{12}}, \bibinfo{pages}{163--167} (\bibinfo{year}{2019}).
\newblock \urlprefix\url{https://www.nature.com/articles/s41561-019-0310-1}.
\newblock \bibinfo{note}{Publisher: Nature Publishing Group}.

\bibitem{goldblatt_earths_2021}
\bibinfo{author}{Goldblatt, C.}, \bibinfo{author}{McDonald, V.~L.} \& \bibinfo{author}{McCusker, K.~E.}
\newblock \bibinfo{title}{Earth's long-term climate stabilized by clouds}.
\newblock \emph{\bibinfo{journal}{Nature Geoscience}} \textbf{\bibinfo{volume}{14}}, \bibinfo{pages}{143--150} (\bibinfo{year}{2021}).
\newblock \urlprefix\url{https://ui.adsabs.harvard.edu/abs/2021NatGe..14..143G}.
\newblock \bibinfo{note}{ADS Bibcode: 2021NatGe..14..143G}.

\bibitem{leconte_increased_2013}
\bibinfo{author}{Leconte, J.}, \bibinfo{author}{Forget, F.}, \bibinfo{author}{Charnay, B.}, \bibinfo{author}{Wordsworth, R.} \& \bibinfo{author}{Pottier, A.}
\newblock \bibinfo{title}{Increased insolation threshold for runaway greenhouse processes on {Earth}-like planets}.
\newblock \emph{\bibinfo{journal}{Nature}} \textbf{\bibinfo{volume}{504}}, \bibinfo{pages}{268--271} (\bibinfo{year}{2013}).
\newblock \urlprefix\url{https://ui.adsabs.harvard.edu/abs/2013Natur.504..268L}.
\newblock \bibinfo{note}{ADS Bibcode: 2013Natur.504..268L}.

\bibitem{yan_cloud_2022}
\bibinfo{author}{Yan, M.}, \bibinfo{author}{Yang, J.}, \bibinfo{author}{Zhang, Y.} \& \bibinfo{author}{Huang, H.}
\newblock \bibinfo{title}{Cloud {Feedback} on {Earth}'s {Long}-{Term} {Climate} {Simulated} by a {Near}-{Global} {Cloud}-{Permitting} {Model}}.
\newblock \emph{\bibinfo{journal}{Geophysical Research Letters}} \textbf{\bibinfo{volume}{49}}, \bibinfo{pages}{e2022GL100152} (\bibinfo{year}{2022}).
\newblock \urlprefix\url{https://onlinelibrary.wiley.com/doi/abs/10.1029/2022GL100152}.
\newblock \bibinfo{note}{\_eprint: https://onlinelibrary.wiley.com/doi/pdf/10.1029/2022GL100152}.

\bibitem{charnay_3d_2015}
\bibinfo{author}{Charnay, B.}, \bibinfo{author}{Meadows, V.}, \bibinfo{author}{Misra, A.}, \bibinfo{author}{Leconte, J.} \& \bibinfo{author}{Arney, G.}
\newblock \bibinfo{title}{{3D} {MODELING} {OF} {GJ1214b}'s {ATMOSPHERE}: {FORMATION} {OF} {INHOMOGENEOUS} {HIGH} {CLOUDS} {AND} {OBSERVATIONAL} {IMPLICATIONS}}.
\newblock \emph{\bibinfo{journal}{The Astrophysical Journal}} \textbf{\bibinfo{volume}{813}}, \bibinfo{pages}{L1} (\bibinfo{year}{2015}).
\newblock \urlprefix\url{https://iopscience.iop.org/article/10.1088/2041-8205/813/1/L1}.

\bibitem{charnay_3d_2015-1}
\bibinfo{author}{Charnay, B.}, \bibinfo{author}{Meadows, V.} \& \bibinfo{author}{Leconte, J.}
\newblock \bibinfo{title}{{3D} {MODELING} {OF} {GJ1214B}'{S} {ATMOSPHERE}: {VERTICAL} {MIXING} {DRIVEN} {BY} {AN} {ANTI}-{HADLEY} {CIRCULATION}}.
\newblock \emph{\bibinfo{journal}{The Astrophysical Journal}} \textbf{\bibinfo{volume}{813}}, \bibinfo{pages}{15} (\bibinfo{year}{2015}).
\newblock \urlprefix\url{https://doi.org/10.1088/0004-637x/813/1/15}.
\newblock \bibinfo{note}{Publisher: American Astronomical Society}.

\bibitem{bardet_joint_2022}
\bibinfo{author}{Bardet, D.}, \bibinfo{author}{Spiga, A.} \& \bibinfo{author}{Guerlet, S.}
\newblock \bibinfo{title}{Joint evolution of equatorial oscillation and interhemispheric circulation in {Saturn}'s stratosphere}.
\newblock \emph{\bibinfo{journal}{Nature Astronomy}} \textbf{\bibinfo{volume}{6}}, \bibinfo{pages}{804--811} (\bibinfo{year}{2022}).
\newblock \urlprefix\url{https://ui.adsabs.harvard.edu/abs/2022NatAs...6..804B}.
\newblock \bibinfo{note}{ADS Bibcode: 2022NatAs...6..804B}.

\bibitem{turbet_water_2023}
\bibinfo{author}{Turbet, M.} \emph{et~al.}
\newblock \bibinfo{title}{Water {Condensation} {Zones} around {Main} {Sequence} {Stars}} (\bibinfo{year}{2023}).
\newblock \urlprefix\url{http://arxiv.org/abs/2308.15110}.
\newblock \bibinfo{note}{ArXiv:2308.15110 [astro-ph, physics:physics]}.

\bibitem{milcareck_radiative-convective_2024}
\bibinfo{author}{Milcareck, G.} \emph{et~al.}
\newblock \bibinfo{title}{Radiative-convective models of the atmospheres of {Uranus} and {Neptune}: {Heating} sources and seasonal effects}.
\newblock \emph{\bibinfo{journal}{Astronomy \& Astrophysics}}  (\bibinfo{year}{2024}).
\newblock \urlprefix\url{https://www.aanda.org/articles/aa/abs/forth/aa48987-23/aa48987-23.html}.
\newblock \bibinfo{note}{Publisher: EDP Sciences}.

\bibitem{boissinot_global_2024}
\bibinfo{author}{Boissinot, A.}, \bibinfo{author}{Spiga, A.}, \bibinfo{author}{Guerlet, S.}, \bibinfo{author}{Cabanes, S.} \& \bibinfo{author}{Bardet, D.}
\newblock \bibinfo{title}{Global climate modeling of the {Jupiter} troposphere and effect of dry and moist convection on jets}.
\newblock \emph{\bibinfo{journal}{Astronomy \& Astrophysics}}  (\bibinfo{year}{2024}).
\newblock \urlprefix\url{https://www.aanda.org/articles/aa/abs/forth/aa45220-22/aa45220-22.html}.
\newblock \bibinfo{note}{Publisher: EDP Sciences}.

\bibitem{dubos_dynamico-10_2015}
\bibinfo{author}{Dubos, T.} \emph{et~al.}
\newblock \bibinfo{title}{{DYNAMICO}-1.0, an icosahedral hydrostatic dynamical core designed for consistency and versatility}.
\newblock \emph{\bibinfo{journal}{Geoscientific Model Development}} \textbf{\bibinfo{volume}{8}}, \bibinfo{pages}{3131--3150} (\bibinfo{year}{2015}).
\newblock \urlprefix\url{https://gmd.copernicus.org/articles/8/3131/2015/}.

\bibitem{meurdesoif_xios_2020}
\bibinfo{author}{Meurdesoif, Y.}
\newblock \bibinfo{title}{{XIOS}: current developments and roadmap} (\bibinfo{year}{2020}).
\newblock \urlprefix\url{http://forge.ipsl.jussieu.fr/ioserver/raw-attachment/wiki/WikiStart/XIOS-ROADMAP-15102020.pdf}.

\bibitem{goody_correlated-k_1989}
\bibinfo{author}{Goody, R.}, \bibinfo{author}{West, R.}, \bibinfo{author}{Chen, L.} \& \bibinfo{author}{Crisp, D.}
\newblock \bibinfo{title}{The correlated-k method for radiation calculations in nonhomogeneous atmospheres}.
\newblock \emph{\bibinfo{journal}{Journal of Quantitative Spectroscopy and Radiative Transfer}} \textbf{\bibinfo{volume}{42}}, \bibinfo{pages}{539--550} (\bibinfo{year}{1989}).
\newblock \urlprefix\url{https://www.sciencedirect.com/science/article/pii/0022407389900447}.

\bibitem{toon_rapid_1989}
\bibinfo{author}{Toon, O.~B.}, \bibinfo{author}{McKay, C.~P.}, \bibinfo{author}{Ackerman, T.~P.} \& \bibinfo{author}{Santhanam, K.}
\newblock \bibinfo{title}{Rapid calculation of radiative heating rates and photodissociation rates in inhomogeneous multiple scattering atmospheres}.
\newblock \emph{\bibinfo{journal}{Journal of Geophysical Research}} \textbf{\bibinfo{volume}{94}}, \bibinfo{pages}{16287--16301} (\bibinfo{year}{1989}).
\newblock \urlprefix\url{https://ui.adsabs.harvard.edu/abs/1989JGR....9416287T}.
\newblock \bibinfo{note}{ADS Bibcode: 1989JGR....9416287T}.

\bibitem{hourdin_meteorological_1993}
\bibinfo{author}{Hourdin, F.}, \bibinfo{author}{Le~van, P.}, \bibinfo{author}{Forget, F.} \& \bibinfo{author}{Talagrand, O.}
\newblock \bibinfo{title}{Meteorological {Variability} and the {Annual} {Surface} {Pressure} {Cycle} on {Mars}.}
\newblock \emph{\bibinfo{journal}{Journal of the Atmospheric Sciences}} \textbf{\bibinfo{volume}{50}}, \bibinfo{pages}{3625--3640} (\bibinfo{year}{1993}).
\newblock \urlprefix\url{https://ui.adsabs.harvard.edu/abs/1993JAtS...50.3625H}.
\newblock \bibinfo{note}{Publisher: AMS ADS Bibcode: 1993JAtS...50.3625H}.

\bibitem{mellor_development_1982}
\bibinfo{author}{Mellor, G.~L.} \& \bibinfo{author}{Yamada, T.}
\newblock \bibinfo{title}{Development of a turbulence closure model for geophysical fluid problems}.
\newblock \emph{\bibinfo{journal}{Reviews of Geophysics}} \textbf{\bibinfo{volume}{20}}, \bibinfo{pages}{851--875} (\bibinfo{year}{1982}).
\newblock \urlprefix\url{https://onlinelibrary.wiley.com/doi/abs/10.1029/RG020i004p00851}.
\newblock \bibinfo{note}{\_eprint: https://onlinelibrary.wiley.com/doi/pdf/10.1029/RG020i004p00851}.

\bibitem{hourdin_use_1999}
\bibinfo{author}{Hourdin, F.} \& \bibinfo{author}{Armengaud, A.}
\newblock \bibinfo{title}{The {Use} of {Finite}-{Volume} {Methods} for {Atmospheric} {Advection} of {Trace} {Species}. {Part} {I}: {Test} of {Various} {Formulations} in a {General} {Circulation} {Model}}.
\newblock \emph{\bibinfo{journal}{Monthly Weather Review}} \textbf{\bibinfo{volume}{127}}, \bibinfo{pages}{822--837} (\bibinfo{year}{1999}).
\newblock \urlprefix\url{https://journals.ametsoc.org/view/journals/mwre/127/5/1520-0493_1999_127_0822_tuofvm_2.0.co_2.xml}.
\newblock \bibinfo{note}{Publisher: American Meteorological Society Section: Monthly Weather Review}.

\bibitem{fuchs_mechanics_1965}
\bibinfo{author}{Fuchs, N.~A.}, \bibinfo{author}{Daisley, R.~E.}, \bibinfo{author}{Fuchs, M.}, \bibinfo{author}{Davies, C.~N.} \& \bibinfo{author}{Straumanis, M.~E.}
\newblock \bibinfo{title}{The {Mechanics} of {Aerosols}}.
\newblock \emph{\bibinfo{journal}{Physics Today}} \textbf{\bibinfo{volume}{18}}, \bibinfo{pages}{73} (\bibinfo{year}{1965}).
\newblock \urlprefix\url{https://ui.adsabs.harvard.edu/abs/1965PhT....18d..73F}.
\newblock \bibinfo{note}{ADS Bibcode: 1965PhT....18d..73F}.

\bibitem{rosner_transport_1986}
\bibinfo{author}{Rosner, D.~E.}
\newblock \emph{\bibinfo{title}{Transport processes in chemically reacting flow systems}} Butterworths series in chemical engineering (\bibinfo{publisher}{Butterworths}, \bibinfo{address}{Boston}, \bibinfo{year}{1986}).

\bibitem{petersen_properties_1970}
\bibinfo{author}{Petersen, H.}
\newblock \bibinfo{title}{The properties of helium: {Density}, specific heats, viscosity, and thermal conductivity at pressures from 1 to 100 bar and from room temperature to about 1800 {K}}  (\bibinfo{year}{1970}).
\newblock \urlprefix\url{https://www.semanticscholar.org/paper/The-properties-of-helium%3A-Density%2C-specific-heats%2C-Petersen/e1fef49cedeed4fe7296f566e20680f74b3dc8a2}.

\bibitem{sengers_representative_1984}
\bibinfo{author}{Sengers, J.~V.} \& \bibinfo{author}{Kamgar-Parsi, B.}
\newblock \bibinfo{title}{Representative {Equations} for the {Viscosity} of {Water} {Substance}}.
\newblock \emph{\bibinfo{journal}{Journal of Physical and Chemical Reference Data}} \textbf{\bibinfo{volume}{13}}, \bibinfo{pages}{185--205} (\bibinfo{year}{1984}).
\newblock \urlprefix\url{https://ui.adsabs.harvard.edu/abs/1984JPCRD..13..185S}.
\newblock \bibinfo{note}{ADS Bibcode: 1984JPCRD..13..185S}.

\bibitem{blain_1d_2021}
\bibinfo{author}{Blain, D.}, \bibinfo{author}{Charnay, B.} \& \bibinfo{author}{Bézard, B.}
\newblock \bibinfo{title}{{1D} atmospheric study of the temperate sub-{Neptune} {K2}-18b}.
\newblock \emph{\bibinfo{journal}{Astronomy \& Astrophysics}} \textbf{\bibinfo{volume}{646}}, \bibinfo{pages}{A15} (\bibinfo{year}{2021}).
\newblock \urlprefix\url{https://www.aanda.org/articles/aa/abs/2021/02/aa39072-20/aa39072-20.html}.
\newblock \bibinfo{note}{Publisher: EDP Sciences}.

\bibitem{leconte_spectral_2021}
\bibinfo{author}{Leconte, J.}
\newblock \bibinfo{title}{Spectral binning of precomputed correlated- \textit{k} coefficients}.
\newblock \emph{\bibinfo{journal}{Astronomy \& Astrophysics}} \textbf{\bibinfo{volume}{645}}, \bibinfo{pages}{A20} (\bibinfo{year}{2021}).
\newblock \urlprefix\url{https://www.aanda.org/10.1051/0004-6361/202039040}.

\bibitem{cooper_modeling_2003}
\bibinfo{author}{Cooper, C.~S.}, \bibinfo{author}{Sudarsky, D.}, \bibinfo{author}{Milsom, J.~A.}, \bibinfo{author}{Lunine, J.~I.} \& \bibinfo{author}{Burrows, A.}
\newblock \bibinfo{title}{Modeling the {Formation} of {Clouds} in {Brown} {Dwarf} {Atmospheres}}.
\newblock \emph{\bibinfo{journal}{The Astrophysical Journal}} \textbf{\bibinfo{volume}{586}}, \bibinfo{pages}{1320--1337} (\bibinfo{year}{2003}).
\newblock \urlprefix\url{https://ui.adsabs.harvard.edu/abs/2003ApJ...586.1320C}.
\newblock \bibinfo{note}{Publisher: IOP ADS Bibcode: 2003ApJ...586.1320C}.

\bibitem{bardet_global_2021}
\bibinfo{author}{Bardet, D.} \emph{et~al.}
\newblock \bibinfo{title}{Global climate modeling of {Saturn}'s atmosphere. {Part} {IV}: {Stratospheric} equatorial oscillation}.
\newblock \emph{\bibinfo{journal}{Icarus}} \textbf{\bibinfo{volume}{354}}, \bibinfo{pages}{114042} (\bibinfo{year}{2021}).
\newblock \urlprefix\url{https://ui.adsabs.harvard.edu/abs/2021Icar..35414042B}.
\newblock \bibinfo{note}{ADS Bibcode: 2021Icar..35414042B}.

\bibitem{cowan_light_2013}
\bibinfo{author}{Cowan, N.~B.}, \bibinfo{author}{Fuentes, P.~A.} \& \bibinfo{author}{Haggard, H.~M.}
\newblock \bibinfo{title}{Light curves of stars and exoplanets: estimating inclination, obliquity and albedo}.
\newblock \emph{\bibinfo{journal}{Monthly Notices of the Royal Astronomical Society}} \textbf{\bibinfo{volume}{434}}, \bibinfo{pages}{2465--2479} (\bibinfo{year}{2013}).
\newblock \urlprefix\url{https://ui.adsabs.harvard.edu/abs/2013MNRAS.434.2465C}.
\newblock \bibinfo{note}{ADS Bibcode: 2013MNRAS.434.2465C}.

\bibitem{burrows_theory_2001}
\bibinfo{author}{Burrows, A.}, \bibinfo{author}{Hubbard, W.~B.}, \bibinfo{author}{Lunine, J.~I.} \& \bibinfo{author}{Liebert, J.}
\newblock \bibinfo{title}{The {Theory} of {Brown} {Dwarfs} and {Extrasolar} {Giant} {Planets}}.
\newblock \emph{\bibinfo{journal}{Reviews of Modern Physics}} \textbf{\bibinfo{volume}{73}}, \bibinfo{pages}{719--765} (\bibinfo{year}{2001}).
\newblock \urlprefix\url{http://arxiv.org/abs/astro-ph/0103383}.
\newblock \bibinfo{note}{ArXiv:astro-ph/0103383}.

\bibitem{kiladis_convectively_2009}
\bibinfo{author}{Kiladis, G.~N.}, \bibinfo{author}{Wheeler, M.~C.}, \bibinfo{author}{Haertel, P.~T.}, \bibinfo{author}{Straub, K.~H.} \& \bibinfo{author}{Roundy, P.~E.}
\newblock \bibinfo{title}{Convectively coupled equatorial waves}.
\newblock \emph{\bibinfo{journal}{Reviews of Geophysics}} \textbf{\bibinfo{volume}{47}}, \bibinfo{pages}{RG2003} (\bibinfo{year}{2009}).
\newblock \urlprefix\url{https://ui.adsabs.harvard.edu/abs/2009RvGeo..47.2003K}.
\newblock \bibinfo{note}{ADS Bibcode: 2009RvGeo..47.2003K}.

\bibitem{maury_presence_2014}
\bibinfo{author}{Maury, P.} \& \bibinfo{author}{Lott, F.}
\newblock \bibinfo{title}{On the presence of equatorial waves in the lower stratosphere of a general circulation model}.
\newblock \emph{\bibinfo{journal}{Atmospheric Chemistry and Physics}} \textbf{\bibinfo{volume}{14}}, \bibinfo{pages}{1869--1880} (\bibinfo{year}{2014}).
\newblock \urlprefix\url{https://acp.copernicus.org/articles/14/1869/2014/acp-14-1869-2014.html}.
\newblock \bibinfo{note}{Publisher: Copernicus GmbH}.

\bibitem{teinturier_clouds_2025}
\bibinfo{author}{Teinturier, L.}, \bibinfo{author}{Charnay, B.}, \bibinfo{author}{Spiga, A.} \& \bibinfo{author}{Bézard, B.}
\newblock \bibinfo{title}{Clouds as the driver of variability and colour changes in brown dwarf atmospheres}  (\bibinfo{year}{2025}).
\newblock \urlprefix\url{https://zenodo.org/records/15777433?token=eyJhbGciOiJIUzUxMiJ9.eyJpZCI6IjkyMzZhZjc5LTQyM2UtNGJiMi04NWFjLTE4NTY2YTMwMmE3YiIsImRhdGEiOnt9LCJyYW5kb20iOiJhNDdkZDY0MTlmNTc3YmJmZjAxODJlYTViMzVmNzcyOSJ9.9TE8W8YCEpvXBwS0WlLbrsrimYj2N_-eAu67cwhgdN923w2qSdwmXYP76TvDxNztgXdA1qKvFwtUaaOEEUFgCw}.
\newblock \bibinfo{note}{Publisher: Zenodo}.

\bibitem{best_ultracoolsheet_2024}
\bibinfo{author}{Best, W. M.~J.} \emph{et~al.}
\newblock \bibinfo{title}{The {UltracoolSheet}: {Photometry}, {Astrometry}, {Spectroscopy}, and {Multiplicity} for 4000+ {Ultracool} {Dwarfs} and {Imaged} {Exoplanets}} (\bibinfo{year}{2024}).
\newblock \urlprefix\url{https://zenodo.org/records/13993077}.

\end{thebibliography}

\end{document}